\documentclass[english, 12 pt]{article}
\usepackage[T1]{fontenc}
\usepackage[latin9]{inputenc}
\usepackage{graphicx}
\usepackage{babel}
\usepackage{url}
\usepackage{multicol}
\usepackage{amsmath}
\usepackage{rotating}           
\usepackage{color}

\makeatletter

\newcommand{\fig}[1]{Figure(\ref{#1})}

\newcommand{\eq}[1]{Eq.(\ref{#1})}
\newcommand{\eqs}[1]{Eqs.(\ref{#1})}

\newcommand{\se}[1]{Sec.(\ref{#1})}

\newcommand{\la}[1]{ \label{#1}}
\renewcommand{\a}{\alpha}
\renewcommand{\b}{\beta}
\renewcommand{\d}{\delta}
\newcommand{\e}{\epsilon}

\renewcommand{\o}{\omega}
\renewcommand{\r}{\mathbf{r}}
\newcommand{\R}{\mathbf{R}}
\newcommand{\p}{\mathbf{p}}
\newcommand{\q}{\mathbf{q}}
\newcommand{\s}{\sigma}

\newcommand{\bsubs}{\begin{subequations}}
\newcommand{\esubs}{\end{subequations}}

\providecommand{\ba}{\begin{align}}
\providecommand{\ea}{\end{align}}

\renewcommand{\S}{\Sigma}
\usepackage{multirow}
\newcommand{\be}{\begin{equation}}
\newcommand{\ee}{\end{equation}}

\newcommand{\bea}{\begin{eqnarray}}
\newcommand{\eea}{\end{eqnarray}}

\begin{document}

\title{ Entropy is in Flux  \\
}

 \author{ Leo P. Kadanoff\footnote{e-mail:  lkadanoff@gmail.com}~
 \\
\\
 The James Franck Institute\\
The University of Chicago
\\ Chicago, IL USA 60637
\\ 
and \\
The Perimeter Institute,\\
Waterloo, Ontario, Canada N2L 2Y5\\
\\}

\maketitle

\begin{abstract}
The science of thermodynamics was put together in the Nineteenth Century  to describe large systems in equilibrium. One part of thermodynamics defines entropy for equilibrium systems and demands an ever-increasing entropy  for non-equilibrium ones.   Since thermodynamics does not define entropy out of equilibrium, pure thermodynamics cannot follow the details of how this increase occurs.      However, starting with the work of Ludwig Boltzmann in 1872, and continuing to the present day, various models of non-equilibrium behavior have been put together with the specific aim of generalizing the concept of entropy to non-equilibrium situations. This kind of entropy has been termed {\em kinetic entropy} to distinguish it from the thermodynamic variety.  Knowledge of kinetic entropy started from Boltzmann's insight about his equation for the time dependence of gaseous systems.   In this paper, his result is stated as a definition of kinetic entropy in terms of a local equation for the entropy density.  This definition is then applied to Landau's theory of the Fermi  liquid thereby giving the kinetic entropy within that theory. 

The dynamics of many condensed matter system including Fermi liquids, low temperature superfluids, and ordinary metals lend themselves to the definition of kinetic entropy. In fact, entropy has been defined and used for a wide variety of situations in which a condensed matter system has been allowed to relax for a sufficient period so that the very most rapid fluctuations have been ironed out. One of the broadest applications of non-equilibrium analysis considers quantum degenerate systems using Martin-Schwinger Green's functions\cite{MS} as generalized of Wigner functions, $g^<(\p,\o,\R,T)$ and $g^>(\p,\o,\R,T)$.   This paper describes once again these how the quantum kinetic equations for these functions give locally defined conservation laws for mass momentum and energy.  In local thermodynamic equilibrium, this kinetic theory enables a reasonable local definition of entropy density.   However, when the system is outside of local equilibrium, this definition fails.  It is speculated that quantum entanglement is the source of this failure.

\end{abstract}

\newpage

\tableofcontents
\newpage
\renewcommand{\theequation}{1-\arabic{equation}}
\setcounter{equation}{0}
\section{"Taking Thermodynamics too Seriously"}
Professional philosophers of physics like sharp definitions; theoretical physicists tend to be more loose.  These facts should not be a surprise to anyone.  Philosophers want to bring out in a precise and careful manner the actual assumptions and content that form the basis for physical theorizing and physical theories.   In the practice of ``normal science\cite{Kuhn},'' scientists are trying to push back the borders of their domains by extending the range of application of their theories.   Thus, they would like some flexibility in their basic definitions so that they can change their content without too much intellectual pain.

These contrasting tendencies have long conflicted in making definitions of entropy.   It is not accidental that the very first use of the word ``entropy'' in English, by George Tait, was prefaced by the statement that  he was going to {\em use use the word in the opposite sense from its German inventor, Rudolf Clasius}\cite{Oxford}. 

This paper was occasioned by thinking about a work\cite{Callender} by the philosopher Craig Callender entitled ``Taking Thermodynamics Too Seriously.'' His basic point was that one should not apply thermodynamic concepts to all kinds of things, but one should stick to the original area of application of thermodynamics\footnote{My difference with Callender is precisely that he takes thermodynamics too seriously. Following the wording and emphasis of Callender I would say that ``The problem is [...] thinking that one is explaining the behavior of {\em individual real systems} by '' appealing to the laws of thermodynamics. Indeed, thermodynamics applies to nothing in this world but instead to the limiting case implied by the words ``the thermodynamic limit''.}.  This area was carefully defined by J. Willard Gibbs in about 1875 in his treatise\cite{GibbsThermo}.  Gibbs described the situation in which thermodynamics laws would hold, the {\em thermodynamic limit,} as one in which one would have a homogeneous system, tending to be infinite in extent, and having been left in a constant external environment for a time tending to be infinitely long.  In this view, any effect of finite size, transient behavior, or observable non-uniformity in structure would push the object under study out of the realm of thermodynamics.  The homogeneity of the system is important since entropy should be an extensive quantity, namely one which grows  linearly with size as the system grows larger.  

 Sadi Carnot's\cite{Carnot}  and Rudolf  Clausius'\cite{Clausius}   thermodynamic analysis provided the original source of this entropy concept.  It is reasonable to call the entropy that comes from their ideas and Gibbs', {\em thermodynamic entropy.}  However, beginning with Boltzmann the concept of entropy has been extended far beyond its original boundaries, so that at present the word ``entropy'' appears in the consideration of black holes, small-$N$ dynamical systems, entanglement\cite{Gilder}, and information theory.  Indeed perhaps, some authors extend this concept beyond recognition.  This paper traces looks at the simplest and most traditional extension, that of {\em kinetic entropy}.  

It looks into three separate contexts: first that of Boltzmann\cite{Boltzmann1872}, then that of Landau\cite{L3}, and finally the one derived from the modern field theory of degenerate quantum systems\cite{Schwinger,KB}. 

\subsection{An alternative definition of entropy}\la{alternative}
In my view as a theoretical physicist, the essential question is not one of precise and careful definition.  Many such definitions are possible and plausible.  Instead one should ask a question which has a answer, for example, {\em Is it possible to define, calculate,  {\em and use} an entropy for a broad class of non-equilibrium systems. } For a working scientist, it is important that the definition be useful for gaining an understanding of the workings of the world, and also for constructing further scientific knowledge.  Nonetheless let me try to give a careful definition of entropy  alternative to that of Callender and the thermodynamicists.  I start:  

\begin{quote}
A dynamical system is a set of equations describing the state of a idealized system aiming to emulate the dynamics of matter.  

The system has a ``state function'' $f(\R, T)$ which describes its state in the neighborhood at the space point, $\R$ at time $T$.   This function may have many components and may depend upon many subsidiary variables.  This state function obeys an autonomous differential equation.  Here ``autonomous'' means that the form of the differential equations depends upon neither $\R$ nor $T$ but only upon $f$

To define kinetic entropy, we imagine three functions of the state function: two scalar,  the entropy density $\rho_s(\R,T)$ and a collision term RHT$_s(\R,T)$, and a spatial vector $\mathbf{j}_s(\R,T)$, all of  which have no explicit space or  time dependence.  They vary in space and time because they each depend upon the values of $f(\R,T)$ in the neighborhood of the space point, $\R$, and they are respectable, reasonably smooth functions of $f$.  

They are related by the equation
\be \la{DefineEntropy} 
\partial_T ~ \rho_s(\R,T) +\nabla_\R \cdot  \mathbf{j}_s(\R,T) 
= \text{RHT} _s(\R,T)~\ge~0
\ee
where the right hand term (RHT)$_s$  is the descendant of Boltzmann's collision term.  It is always positive\footnote{For appropriate functional forms of the distribution function, positive, finite, and sufficiently smooth.  } except when $f(\R,T)$ for all $\R$ and some finite interval of times get a very special, time-independent form, called an equilibrium distribution. One can call the entropy thus defined {\em kinetic entropy} to distinguish it from thermodynamic entropy. 

Under these circumstances,  the $S(T)=\int d\R~ \rho_s(\R,T)$  is an entropy  which describes the non-equilibrium state of the system. 

\end{quote}

 Using this approach, Boltzmann\cite{Boltzmann1872} defined entropy for a dilute gas using \eq{DefineEntropy}(See \se{Boltzmann}.);  while Landau\cite{L3} set up a quasiparticle dynamic for some low temperature fermion systems that enable the definition of entropy (See \se{Landau}.).  On the other hand, we do not know whether field theoretic Green's function methods\cite{Schwinger,KB,Keldysh} enable the definition of entropy in non-equlibrium situations.

If on the other hand, if it is impossible to construct an equation of the form of \eq{DefineEntropy}, but the best one can do  with smooth and local functions is perhaps something like
\bea  \la{DefineNoEntropy} 
&~&\partial_T \rho_s(\R,T) + \nabla_\R  \mathbf{j}_s(\R,T)+\sum_jA_j(\R,T)\partial_T B_j(\R,T)
\nonumber \\
&~&~~~~~~~~+\sum_j \mathbf{C}_j(\R,T) \cdot \nabla_\R D_j(\R,T)  = \text{RHT} _s(\R,T) \ge~0,
 \eea
then the concept of kinetic entropy is undefined away from the regions of the $f$-space in which all the  summation-terms are zero.  

It should be noted that the constraints of smoothness and locality are crucial.   It is likely that, dropping these constraints, one could define an entropy concept in which there was a global, increasing, entropy that might form a fractal web through the space  entire system.    Such an entropy might apply to entangled\cite{Gilder} quantum systems, but would be vastly different from Boltzmann's kinetic entropy.\footnote{In a chaotic classical system we have a large number of constants of the motion which have level sets in the form of  a complex web through the entire coordinate space.  When we talk about ``conserved quantities,'' we do not usually include these very complex objects. Perhaps similarly the existence of entanglement entropy might preclude the existence of Boltzmann-like kinetic entropy.    }

\subsection{Outline of paper}

The next chapter, \se{Boltzmann}, goes back to Boltzmann who showed how rarified gases could be described with the use of three conceptually different kinds of entropy.  They are the thermodynamic entropy \`{a} la Clausius, the kinetic entropy that describes the gas' relaxation to thermodynamic equilibrium, and a statistical entropy that is expected to be the logarithm of the number of micro-states subsumed by a single macroscopic configuration. All three entropies can be defined to have the same numerical value in thermodynamic equilibrium.   However, Boltzmann's three-fold extension was obtained at the cost of an assumption that only applies in a very special limit:  one in which the interaction among the gas molecules be so weak as to provide no direct effect upon the gas-state thermodynamics.   

In the Landau theory of a Fermi liquid\cite{L3,L4,L5}, described in  \se{Landau}, all interactions are exactly included but one assumes that  the system has the smoothness and homogeneity of a fluid.  The validity of the Landau theory also relies upon low temperatures to force the system into a dynamics based upon  excitations that can be described as {\em quasiparticles}\cite{Bogoliubov,Kaganov}.  In this case also,  three kinds of entropy can be calculated and agree with one another in appropriate limits. Once again, as in Gibbs description of thermodynamics, one must make assumptions about the system being sufficiently large and having had sufficient time-left-alone to relax the most complex correlations.

  Field theory\cite{MS,Schwinger,Keldysh} provides a more general quantum theory that can be shaped into a quantum version of the Boltzmann kinetic equation\cite{KB}. This generalized Boltzmann equation includes both frequencies and momenta in its description of excitations.  I expected this approach to provide a more general derivation of the three three kinds of entropy.  So far, I have failed to achieve this.  Instead, I have much weaker statement derived from this point of view, namely the derivation of kinetic entropy, \eq{DefineEntropy},  seems to hold only in the limit of local thermodynamic equilibrium.  However,   the approach in \se{Green} seems to give in general \eq{DefineNoEntropy} rather than \eq{DefineEntropy}.  Therefore the first two approaches (Boltzmann and Landau) are fully consistent with the definition of kinetic entropy but the last may well not be so.  
  
  I would guess that this last result means that the field theoretical model of \se{Green} includes some sort of entanglement\cite{Gilder} in its dynamics,  and that entanglement entropy is not included within the definition, given above,  of kinetic entropy.

\renewcommand{\theequation}{2-\arabic{equation}}
\setcounter{equation}{0}
\section{Boltzmann:  Three Kinds of Entropy} \la{Boltzmann}
\subsection{ Thermodynamic entropy}
The concept of entropy was put together in the nineteenth century by such giants as Carnot, Clausius, Boltzmann, and Gibbs.  They constructed  a beautiful and logically complete theory.  Their work reached one kind of culmination with the 1875 and 1878 publication of J. Willard Gibbs\cite{GibbsThermo}, which described the entire theory of thermodynamics and of the ``thermodynamic limit''. This is the situation which would arise in most macroscopic systems if left alone for long enough and thereby allowed to relax to a steady situation.   Each such isolated system has a property called {\em entropy}.   The entropy of a totality is the sum of the entropies of its individual isolated parts. If such a system is driven out of equilibrium, for example by bringing together two of its parts initially at different temperatures, the system will return to equilibrium, but in the process one observes that its entropy has increased.  It is, of course, natural to ask whether entropy can be defined during the period in which the system is out of equilibrium.  This paper discusses partial answers to this question.

\subsubsection{Maxwell and the basics of flow and collision}
\begin{quotation}
When two molecules come within a certain distance of one another a mutual action occurs between them which may be compared to the collision between two billiard balls.  Each molecule has its course changed and starts on a new path.  James Clark Maxwell.  Theory of Heat\cite{MaxwellHeat}[page 302, original text 1871]
\end{quotation}
James Clark Maxwell was one of the founders of the kinetic theory of gases. His work arose before an atomic theory was generally accepted, but nonetheless he (and his near- contemporary Ludwig Boltzmann) saw a gas as a multitude of very small particles in rapid motion.  These particles each would move in a straight lines, each exhibiting constant momentum, $\p$, and velocity, except that  
\begin{itemize}
\item  they would gradually be deflected by slowly varying forces, like those from gravity and electromagnetic fields
\item and also, they would very suddenly change their direction of motion as they  bumped into one another\footnote{This breakup of the forces in the system into two parts, analyzed differently, follows the classic work of van der Waals\cite{Waals}.}.    
\end{itemize} 
   Maxwell was particularly interested in the transport of energy and momentum through the system, since this transport could be studied in the laboratory\cite{MaxwellDyn}.  For this purpose he constructed what he called transport equations, describing the motion of observable quantities through his gases.  These observables were functions of the momenta, $\p_a$, of the individual gas particles,  with $\a$ being an index defining a particular particle of the gas.   The observable might then be the momentum itself,  $\p_\a$, its position, $\R_\a$,  the particle's kinetic energy, $p_\a^2/(2m)$, or indeed any function of the particle's position and momentum.   The transport equations then described how the observable quantities might change in light of these processes just mentioned.     When he would focus upon a given momentum-value, $\p$, Maxwell could enumerate the number of collisions per unit of time which would have the effect of knocking particles out of that particular momentum-value and equally, he could keep track of the number of particles entering that value in virtue of collisions.   When all this was done, Maxwell could estimate, or in a particularly favorable case, calculate the rate of change of his observables in time.  Thus, he could describe both the dynamical processes involving the transport of energy, particles, or momentum through the gas.   In addition, by looking at situations in which the state of the gas was unchanging,  he could also define the nature of the equilibrium of the system, via the famous Maxwell or Maxwell-Boltzmann distribution.  
    
\subsubsection{Boltzmann's $f(\p,\R,T)$}
Boltzmann took Maxwell's elegant but limited calculations and converted them into a generally formalism for the evaluation of gas properties.   He changed the focus of the calculation from the consideration of particular observable properties of the gas  to a calculation of the rate of change of an average number of particles in a given region with a given momentum, $f(\p,\R,T)$.   
Maxwell's and Boltzmann's original definition\footnote{Boltzmann's original \cite{Boltzmann1872} definition actually used the kinetic energy as the independent variable in his analysis\cite{CC}[page 89].  This is an  not an elegant choice compared to $\p$, and has not been used extensively by later authors.} of this quantity\cite{MaxwellDyn,Boltzmann1872} is that
\be \la{f}
f(\p,\R,T)~d\p ~ d \R
\ee
is the number of particles we will find at time $T$ within the small volume $d \R $ of ordinary space and the small volume $d\p$ of a space describing the possible momenta of the particles.   However, since many of their calculations involved averaging processes,  this definition did not stand up to closer analysis. Eventually, Boltzmann changed his interpretation\cite{Boltzmann1884} to one involving probabilities. Gibbs completed the analysis in his work on statistical mechanics\cite{GibbsStatMech} .  These authors interpreted probabilistic calculations, including ones involving  $f$, by considering the average behavior over many possible similarly prepared systems.  Such a collection of systems eventually came to be called an {\em ensemble}, and this kind of calculation {\em statistical mechanics}. 

Using the ensemble definition of $f$,  we can say that the total, averaged, amount  of observable per unit volume in the region around $\R$ at time $T$  could then be defined as
\be \la{O}
<O(\p,\R)>_{T} =  \int  ~d\p ~ f(\p,\R,T)~ O(\p,\R))
\ee
Here $<.>$ indicates an average over the ensemble.  (Note that we integrate over  possible momenta, $\p$, , while holding the space and time coordinates constant.)
Boltzmann's calculational strategy was then to first write an equation to calculate the distribution function,  $f$, rather than the individual particle positions and momentum, and then use $f$ to calculate the average of observables.

In either view, deterministic or statistical,  Boltzmann's $f$-function exists in six dimensions plus time. ( Indeed, Boltzmann was a quite brave theorist. No complexity fazed him. )

Even the original incorrect, deterministic,  definition of $f$ was an important advance. It permitted Boltzmann to write down his celebrated ``Boltzmann kinetic equation, ``BKE,'' from which he was able to deduce the main properties of gases and to invent statistical mechanics.  I here write down this equation in modern language, so that I may describe its contents.  The equation is
\be
\la{BKE}
\frac{\partial f(\p,\R,T)}{\partial T}+ \{\e(\p,\R,T), f(\p,\R,T)\}= - \s^>(\p,\R,T)f(\p,\R,T) +\s^<(\p,\R,T)
\ee

The BKE describes fully the behavior of low density gases composed of structureless particles interacting via short ranged forces.  It enabled Boltzmann to describe the main dynamical properties of these gases and to construct the basis for a tremendous amount of further work in the dynamics of condensed matter systems.   Further, equations with a similar form have been used to describe the kinetic properties of a wide variety of different condensed matter systems up to the present day\cite{Oettinger}.

Starting from the left-hand side of \eq{BKE}, we see the time derivative of the distribution function $f(\p,\R,T)$.  The next term describes the gradual changes in momentum and position of the particles described by $f$.  This term reflects the formulation of   mechanics described by William Rowen Hamilton\cite{HamiltonMechanics} in 1833. Here, $\e$ is the energy of a single particle as it depends upon particle momentum. It may also include  some space- and time- dependent potentials, as for example electromagnetic potentials.  In that case, it is possible for the energy, $\e$, to dependent upon space, $\R$ and time, $T$.  The bracket $\{.,.\}$ is called a Poisson bracket and is defined by
\be \la{Poisson}
 \{A, B\} =\nabla_\p A \cdot \nabla_\R B -\nabla_\R A \cdot \nabla_\p B ,
\ee
In the BKE of \eq{BKE}, $A$ is the one-particle Hamiltonian and $B$ is $f$,\; and this Poisson bracket describe how $f$ gradually changes with time. The change occurs  because each particle's  position variable, $\R_\a$ changes as it moves with the Hamiltonian velocity\footnote{ \la{footnote}The reader might worry that the signs coming from the Poisson bracket of  BKE are the opposite the signs in \eq{Hamiltonian}.  That is because the variables $\R$ and $\p$ in $f(\p,\R,T) $ are not directly  the position and momentum of any particle.  Instead they are parameters defining which particle momentarily appears with these phase space  coordinates at the instant T.  The definition is 
\be
f(\p,\R,T)= < \sum_\a \delta(\p-\p_a) \delta(\R-\R_a)>.
\nonumber \ee
and <.>  represents an ensemble average.}      
\bsubs \la{Hamiltonian}
\be
\frac{d\R_\a}{dT}= \nabla_\p \e(\p_\a,\R_\a,T)
\ee
while  its momentum is varying in virtue of the force
\be 
\frac{d\p_a}{dT}  = -\nabla_\R \e(\p_a,\R_\a,T).
\ee 
\esubs
The right hand side of the Boltzmann equation describes collision processes.  In the first term, proportional to $f$, the quantity $\s^>(\p,\R,T)$ describes the number of scatterings of particles with momentum $\p$ per unit volume and unit time in the neighborhood of the space-time point $\R,T$.  The minus sign, of course, indicates that this process decreases the number of particles with momentum $\p$.  The remaining term, involving $\s^<(\p,\R,T)$ describes the opposite process, that is   the scattering of particles that then go into the momentum state, $\p$.  Thus, $\s^<$ measures the number of particles going through this process per unit volume and unit time.  Both scattering terms are complicated integrals involving four different momenta, two before the collision and  and two after.  These terms make the BKE into a non-linear integro-differential equation, quadratic in $f(\cdot,\R,T)$.

These scattering terms are most effectively described by setting down the total rate of scattering of all the particles in the system. (See \cite{CC}[pages 88-93].)  The rate at which particles with momentum $\p$ and $\q$ scatter off one another and end up respectively with momenta $\p'$ and $q'$  can be written as a functional of Boltzmann's distribution function, $f$. The total scattering rate is proportional to 
\bsubs \la{ClassicalScattering}
\be \la{psiClassical0}
 \psi = \frac 12 \int ~d\p~d\q~   d\p'~d\q' ~ f(\p)  f(\q)
~\mathcal{P}\left(\begin{array} {ccc} \p &  \rightarrow& \p' \\
\q &  \rightarrow& \q' \\
 \end{array}\right)
\ee
In an appropriate limit of a low density gas with short ranged interactions the collision rate is exactly proportional to this expression, with an appropriate $\mathcal{P}$. (This limit is described as the Grad Limit\cite{Grad}.)     
The factor of one half is inserted for later notational convenience. 

\begin{figure}[h]
\begin{center}
\includegraphics[height=5cm ]{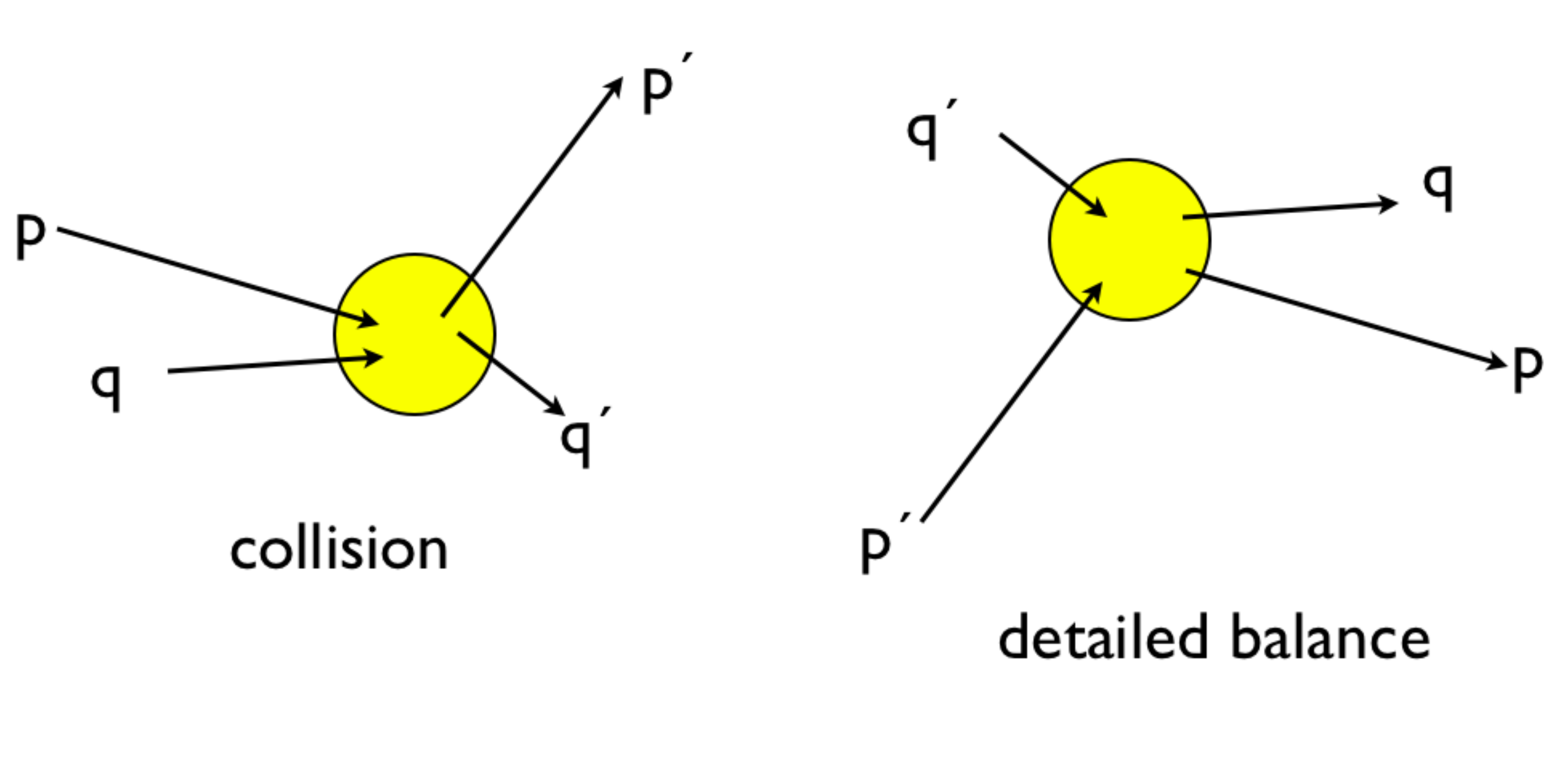}
\end{center}
	\caption{Detailed balance symmetry.}
\label{Balance}
\end{figure}

This process is depicted in \fig{Balance}. The function $~\mathcal{P}$ describes the exact scattering rate  and is derived from a solution of the equations of motion obeyed by the colliding particles.  We shall not specify it in detail here, but only focus upon the symmetries which must be obeyed by this quantity for the Boltzmann equation to make any sense.  In Boltzmann's work $\mathcal{P}$ was specified to describe the result of the classical-mechanical analysis of the two-particle scattering 

This $\psi$ can then we used to find the collision terms in the Boltzmann equation. When $\mathcal{P}$ is properly normalized, we read off the scattering rate $\s^>$   from $\psi$ as 
\bea  \la{s>}
\s^>(\p,\R, T)  
&=&  \int ~d\q~   d\p'~d\q'  ~ f(\q,\R,T)~ ~\mathcal{P}\left(\begin{array} {ccc} \p &  \rightarrow& \q \\
\p' &  \rightarrow& \q' \\
 \end{array}\right) \eea
This same logic gives us an expression for the rate of scattering into the momentum $\p'$:  
\bea \la{s<}
\s^<(\p',\R,T) &=& 
\int d\p~d\q~   ~d\q'  f(\p,\R,t)  f(\q,\R,t)~
 ~\mathcal{P}\left(\begin{array} {ccc} \p &  \rightarrow& \p' \\
\q &  \rightarrow& \q' \\
 \end{array}\right) \nonumber \\
\eea
It will be important for us that the collisions include the conservation of particle number, momentum and energy.  The number conservation is  built into the structure in its demand for having precisely two particle both before and after the collisions. 
 We further demand that $~\mathcal{P}$ be zero unless the conservation rules  are satisfied.  This will occur when $~\mathcal{P}$ is proportional to Dirac delta functions in the form
\bea \la{CollisionConservation}
& ~&\mathcal{P}\left(\begin{array} {ccc} \p &  \rightarrow& \p' \\
\q &  \rightarrow& \q' \\
 \end{array}\right) \  
 =  \mathcal{Q}\left(\begin{array} {ccc} \q &  \rightarrow& \q' \\
\p &  \rightarrow& \p' \\
 \end{array}\right)   \delta(\p+\q-\p'-\q') \quad \quad  \quad \quad \quad \quad  \nonumber \\
 &~& \quad \quad \quad \quad   \quad
 \delta(\e(\p,\R,T)+\e(\q,\R,T)-\e(\p',\R,T)-\e(\q',\R,T)) \nonumber \\
\eea
where $\mathcal{Q}$ depends upon combinations of the momentum variables not fixed by the delta functions.  
\esubs

\subsection{Analysis of the Boltzmann Kinetic Equation (BKE).}
\subsubsection{Symmetries of the BKE}
When we substitute the results of \eq{s>} and \eq{s<} into the expression of \eq{BKE} we have a formulation of the equation called the {\em Boltzmann's kinetic equation}, which will then form the basis of the rest of this paper.  However our formulation is slightly different from the BKE originally used in 1884. The original work had a much more explicit form for the collision function, $\mathcal{Q}$, based upon solving equations of classical mechanics for two particles and a central force.  One advantage of our more generic formulation is that we do not have to look into the details of the scattering calculation.  A subsidiary  advantage of not specifying $\mathcal{Q}$ explicitly is that the BKE thus written applies very broadly.  In fact, it equally applies to  quantum gases in the low density limit, with the previously stated requirement is that the intermolecular forces be short ranged.  In the quantum case, one uses the differential cross-section to specify the scattering rate.    

Since the structure of $\mathcal{Q}$, and not its details, matters to us here, let us outline the needed symmetries . These include
\begin{itemize}
\bsubs
\item Interchange symmetry.  For simplicity we assume that the scattering particles have identical properties so that 
\be \la{InterchangeSymmetry} 
 ~\mathcal{Q}\left(\begin{array} {ccc} \p &  \rightarrow& \p' \\
\q &  \rightarrow& \q' \\
 \end{array}\right) =
 ~\mathcal{Q}\left(\begin{array} {ccc} \q &  \rightarrow& \q' \\
\p &  \rightarrow& \p' \\
 \end{array}\right) 
 \ee
 \item Detailed Balance. Another important symmetry is one in which the primed particles come together, collide, and then the unprimed ones come off.  If scattering rates for this kind of event (see \fig{Balance}) are the same as the rate for the corresponding direct process
\be \la{DetailedBalance}
\mathcal{Q}\left(\begin{array} {ccc} \p &  \rightarrow& \p' \\
\q &  \rightarrow& \q' \\
 \end{array}\right)    =
 \mathcal{Q}\left(\begin{array} {ccc} \p' &  \rightarrow& \p \\
\q' &  \rightarrow& \q \\
 \end{array}\right)    
\ee
the system is said to have a {\em detailed balance} symmetry. 
\item Positivity. $\mathcal{Q}$ is always positive or zero.
\item Galilean symmetry.  The scattering rate will be the same for an observer at rest as for one in motion with velocity $\mathbf{u}$.  If the mass of both scattering particles is $m$ then
\be 
 ~\mathcal{Q}\left(\begin{array} {ccc} \p+m\mathbf{u} &  \rightarrow& \p'+m\mathbf{u} \\
\q+m\mathbf{u} &  \rightarrow& \q'+m\mathbf{u} \\
 \end{array}\right) =
 ~\mathcal{Q}\left(\begin{array} {ccc} \p &  \rightarrow& \p' \\
\q &  \rightarrow& \q' \\
 \end{array}\right) 
\ee
\item Rotational symmetry.  If all the momentum vectors undergo the identical rotation, $\mathcal{Q}$ remains unchanged. 
\item Time reversal symmetry.  This holds in the absence of magnetic fields. It is represented by $(\p,\q) \leftrightarrow (-\p'.-\q')$.   Under this transformation,  $\mathcal{Q}$ remains unchanged. 
\item Parity, Flipping the sign of an odd number of components of all the momenta leaves $\mathcal{Q}$ unchanged.
\esubs 
\end{itemize}

\subsubsection{Boltzmann's conserved observables}
The next step is to write down equations for the time dependence of observables, and thereby trace the route from Boltzmann's kinetic equation back to Maxwell's transport equations.  Multiply the kinetic  equation, \eq{BKE},  by $O(\p,\R)$ and integrate over $\p$  to find 

\bea \la{ChangeOfObservable}
 \frac{d} {dT}<O(\p,\R)>_T &+&\int ~d \p~ \{\e(\p,\R,T), f(\p,\R,T)\} ~ O(\p,\R) 
 \nonumber \\ &=&\int ~ d \p~ O(\p,\R)[ - \s^>(\p,\R,T)f(\p,\R,T) +\s^<(\p,\R,T)]  \nonumber \\
\eea
Note that, for simplicity, we have taken the observable to have no explicit time dependence. The observable, then, only  dependence upon time arises from particles' motion in phase space. 
\bsubs

There are three terms in \eq{ChangeOfObservable}. The first is simply the time rate of change of the observable.  The second, 
\be \la{LHTB}
\text{LHT}_O =  \int ~d \p~ \{ \e, f\} ~ O  =\nabla_\R \cdot \int ~ d\p ~ f~ (\nabla_\p \e) O +\int  d \p~ f~\{ O, \e\} 
 \ee
 comes from the time-dependence of $f$ in virtue of the one-body Hamiltonian $\e$. The divergence term in \eq{LHTB} describes the rate of change produced by the flow of the observable, which then produces a current 
 \be \la{current}  \mathbf{j}_O =  \int d\p ~ f(\p,\R,T)~ (\nabla_p \e(\p,\R,T) ) ~O(\p,\R).
  \ee
  The bracket term on the right of \eq{LHTB} is the intrinsic change in the observable produced by the time derivative of the positions and momenta of particles in the neighborhood of $\R$.  We describe this term as the negative of a {\em generalized force}, $F_O$. Then the entire kinetic equation is the statement
\be \la{ConservationLaws}
d_T <O(\p,\R)>_T +\nabla_\R \cdot \mathbf{j}_O(\R,T) - \mathbf{F}_O(\R,T)=\text{RHT}_O 
\ee
with the right hand terms,  RHT$_O$,  being the collision terms,  
 \be \la{RHT}  \text{RHT}_O =  \int d\p ~ O(\p,\R) [-f(\p,\R,T) \s^>(\p,\R,T)+ \s^<(\p,\R,T)]  \ee
\esubs 

\subsubsection{Collision analysis--variational method}\la{ClassicalVariational}
We follow Boltzmann and look at the observables most significant for understanding the long-time behavior of the system.  These special observables are the conserved quantities, those which are time-independent in an isolated system.  They are the ones which are not changed by collisions terms, so that, for them, RHT$_O=0$   

We could analyze the collision terms quite directly using the symmetries we have just listed. Instead, we employ a method based upon variational analysis.  This method is somewhat too ponderous to be the most elegant method of understanding the BKE.  However, the very same method will come into it own in the analysis of quantum problems, so we introduce it here. 

We consider the total scattering rate as a functional of both the density of scatterers, which will be indicated here as $g^<$ and the density of final states, denoted by $g^>$.  The total scattering rate is then proportional to

\bsubs \la{ClassicalScatteringExtended}
\be \la{psiClassical}
 \psi[g^<,g^>] = \frac 12  \int ~d\p~d\q~   d\p'~d\q' ~  g^<(\p) g^<(\q) g^>(\p') g^>(\q')
~\mathcal{P}\left(\begin{array} {ccc} \p &  \rightarrow& \p' \\
\q &  \rightarrow& \q' \\
 \end{array}\right)
\ee
To get the rate of scattering of a beam of particles with momentum $\p$ we calculate a derivative as
\bea \la{psis>}
 \s^>(\p,\R,T) &=&  
 \left. \frac {\d \psi[g^<,g^>]}{\d g^<(\p)}\right|_{g^<(\cdot)=f(\cdot,\R,T),g^>=1}
\nonumber \\  
 &=&   \int ~d\q~   d\p'~d\q' ~   f(\q,\R,T)
~\mathcal{P}\left(\begin{array} {ccc} \p &  \rightarrow& \p' \\
\q &  \rightarrow& \q' \\
 \end{array}\right) \nonumber \\  
\eea
In precise analogy, to calculate the scattering rate of particles into the beam, we use
\bea \la{psis<}
 \s^<(\p',\R,T) &=&  
 \left. \frac {\d \psi[g]}{\d g^>(\p')}\right|_{g^<(\cdot)=f(\cdot,\R,T),g^>=1}
\nonumber \\  
 &=&   \int~d\p ~d\q~   d\q' ~   f(\p,\R,T) f(\q,\R,T)
~\mathcal{P}\left(\begin{array} {ccc} \p &  \rightarrow& \p' \\
\q &  \rightarrow& \q' \\
 \end{array}\right) \nonumber \\  
\eea
\esubs
This formulation is particularly useful in evaluating the combination
\bsubs \la{VariationalReplacement}
\be  \la{combo}
\chi[O^<,O^>]= \int ~d\p ~O^<(\p)~g^<(\p)~ \frac{\d \psi}{\d g^<(\p)} +
\int~ d\p ~ O^>(p) ~g^>(\p)~ \frac{\d \psi[g]}{\d g^>(\p)}
\ee
which has the effect producing a sum of terms in which each term has a $g^<$ replaced by $O^< g^< $ or in which a $g^>$ replaced by $O^> g^> $, that is
\bea  \la{result}
\chi[O^<,O^>]
&=&
 \frac 12  \int ~d\p~d\q~   d\p'~d\q' ~  g^<(\p) g^<(\q) g^>(\p') g^>(\q')
\nonumber \\ 
&~& [O^<(\p) +O^<(\q) +O^>(\p')+O^>(\q')]
 \nonumber \\ 
&~&\mathcal{P}\left(\begin{array} {ccc} \p &  \rightarrow& \p' \\
\q &  \rightarrow& \q' \\
 \end{array}\right) 
\eea
\esubs
 (We have used the symmetry of \eq{InterchangeSymmetry} to form this result. This symmetry is implicit in our use of the variational derivative to form $s^>$ and $s^<$.)  We describe the analysis in \eqs{VariationalReplacement} as a {\em variational replacement}. We shall see similar replacements in our fully quantum analysis of  \se{QuantumReplacement}

  Specifically, the combination used in calculating the effects of our observables upon the right hand side of the Boltzmann equation is of this form with $O^>(\p)=O(\p,\R)$ and $O^>(\p)=-O(\p,\R)$.  As a result we find that the effect of the observable upon the right hand side of the Boltzmann equation is   
\bea \la{conservation}
 \text{RHT}_O &=&  -\frac12 \int d\p~d\q~ d\p'~d\q'~  [O(\p,\R) +O(\q,\R) -O(\p',\R) -O(\q',\R)]
 \nonumber \\
 & ~&\mathcal{P}\left(\begin{array} {ccc} \p &  \rightarrow& \p' \\
\q &  \rightarrow& \q' \\
 \end{array}\right)    ~   f(\p,\R,T) f(\q,\R,T)
\eea
   We can now simply read off conserved quantities by noting the delta functions (see \eq{CollisionConservation} ) in the collision terms.  

The conservation laws are defined by the values of $O$ that make the integrated right hand terms, as defined by \eq{conservation} equal to zero.   These include the momentum ($O=\p$), angular momentum ($O=\R \times\p$), energy ($O=\e$), and particle number, ($O=1$). 

It is particularly crucial that any approximate kinetic theory include variables that obey these conservation laws, since  the conservation laws drive the low frequency, long wavelength behavior of the system.  For macroscopic observers, like ourselves, this slow variation dominates the immediately accessible behavior of the many-body system.  


\subsubsection{Entropy.}  A very similar approach will work for the derivation of an equation for the entropy density. Boltzmann multiplied the Boltzmann kinetic equation, \eq{BKE}, by an observable that generates the thermodynamic entropy density,  
\be 
O_s(\p,\R,T)=1+ ~\ln (1/f(\p,\R,T))
\ee
and integrated over all momenta. The resulting equation has almost but not quite the form of a conservation law.  It is
\bsubs
\be 
\partial_T \rho_s(\R,T)  +\nabla_\R \cdot j_s(\R,T) =  \text{RHT}_s \ge 0
\ee

This equation describes a continually growing (or constant) entropy  since the right hand side is greater than or equal to zero, and the left hand side describes how the entropy, with density $\rho_s$, 
\be 
\rho_s(\R,T)   =   \int d\p ~ f(\p,\R,T) ~\ln (1/f(\p,\R,T))
\ee
is moved from place to place by the current $\mathbf{j}_s$. 
\be 
 j_s(\R,T)  = \int d\p~ (\nabla_\p \e(\p) ) f(\p,\R,T) ~\ln (1/f(\p,\R,T)) 
\ee
Since only the collisions increase entropy, the generalized force associated with the entropy density is zero.

The crucial analysis is that of the collision term.  We once again use the analysis of \eq{VariationalReplacement}  to get an expression for what we hope to be the entropy production rate in the form
\bea \la{RHT_sc}
RHT_s &=& \frac 12\int ~d\p~d\q~   d\p'~d\q' 
~ \mathcal{P}\left(\begin{array} {ccc} \p &  \rightarrow& \p' \\
\q &  \rightarrow& \q' \\
 \end{array}\right)  
 \nonumber \\
 &~&   [\ln f(\p,\R,T)+~\ln f(\q,\R,T)-~\ln f(\p',\R,T)-~\ln f(\q',\R,T)] 
 \nonumber \\ 
 &~& f(\p,\R,T)  f(\q,\R,T) 
\eea 
\esubs

Following in the footsteps of Boltzmann, we should make two demands of our putative expression for the local entropy production rate: It should be zero for systems in equilibrium and positive for systems out of equilibrium.  The first requirement is automatically satisfied since the equilibrium (Maxwell-Boltzmann) condition upon $f$ is that $\ln f$ be proportional to a sum of the conserved quantities 
\be
\ln f= -\b[ \e(\p-m \mathbf{u}) - \mu]=-\b[ \e(\p)-\p\cdot\mathbf{u} + \frac12 m u^2-\mu]
\ee
so that the term in square brackets in \eq{RHT_sc} exactly vanishes in virtue of the delta functions in $\mathcal{P}$.
When $f(\p,\R,T)$ has its equilibrium form, we say that the system is in {\em local equilibrium}. {\em Local} because our derivation permits the inverse temperature, $\b$, the average particle velocity $\mu$, and the chemical potential, $\mu$ to vary in space and time.  

To complete his argument, Boltzmann found a sufficient, but not necessary condition for this non-negativity, specifically that  $P$ obey the detailed balance condition  of \eq{DetailedBalance},
so that the collision process is equally likely forward as backward in time.  In the case considered here,  this {\em detailed balance} condition is a consequence of time reversal invariance together with the symmetry produced by reversing the direction of all the momentum vectors.    If the balance condition is satisfied, the expression in the collision term of \eq{RHT_sc}, not including the last line,  is anti-symmetric under the interchange  of primed and unprimed variables.  One can make the entire expression symmetric, without changing its values by making the replacement$$
f(\p,\R,T)  f(\q,\R,T) \rightarrow \frac 12 [ f(\p,\R,T)  f(\q,\R,T)-f(\p',\R,T)  f(\q',\R,T) ]
$$
After this replacement,  there are two expressions in \eq{RHT_sc} (the ones in square brackets), each functions of the $f$'s, both of which have a sign that depends upon whether $f(\p)f(\q)$ is greater than or less then  $f(\p')f(\q')$.  Equality occurs only when these are equal, and this equality in turn happens only at local equilibrium.  Therefore this right hand side is positive in all situations except when there is local equilibrium.    

Boltzmann's result is now often called the $H$-theorem because later authors described the result in terms  of  $H$, the negative of the entropy.

\subsection{Two interpretations of $f$}\la{Two interpretations}
We have already noted in writing \eq{f} that in the course of time Boltzmann  $f$ stopped saying that $f$ was the actual number of particles in a given region of phase  space,  and instead  came to interpret $f$ as   an average density of particles in different regions of phase space.


The difference between $f$ defining a number of particles, and $f$ defining a probability is crucial to understanding the true meaning of the Boltzmann equation.    It is the difference between a deterministic and a probabilistic definition of Boltzmann's calculation.    After the original publication of the equation, Boltzmann's contemporaries, especially\cite{CC}[pages 97-102]  Loschmidt and  Poincar\'{e}, pointed out that the Boltzmann equation could not be an exact representation of any situation in classical mechanics.  Classical mechanics, they said, was reversible.  If you reversed the direction of all the velocities in the system, the previous motion unfolded once more, but run backward!    However, the Boltzmann equation exhibited no such behavior.  The left hand side  of \eq{BKE} changes sign with this reversal; the right hand side remains unchanged.  Clearly this equation's behavior cannot reflect an exact property of any single configuration of classical particles.  However, as Boltzmann realized and Howard Grad\cite{Grad} eventually showed, the equation could be describing a probability distribution and an average over similar systems. 

\subsection{Three kinds of entropy}
So far we have followed Boltzmann's development of a theory of kinetic entropy.  This entropy is defined as an extensive quantity (i.e. a sum over locally defined quantities) with the property that it is continually increases until thermodynamic equilibrium is reached, whereupon it attains a constant value.    It is distinguished from the thermodynamic entropy by being defined in out-of-equilibrium situations.   By adding appropriate amounts of the other extensive conserved quantities to the kinetic entropy  one can make the two entropies identical in equilibrium situations.    

Boltzmann also found one more version of the entropy.  In statistical mechanics, one averages over ensembles.    The average has a weight that depends on all the variables in the system and their correlations.    To calculate such an average, one uses a probability density in classical calculations or a density matrix in quantum calculations, both denoted as $\rho$,  and a sum over configurations called a trace. Thus the average of an observable is
$$
<O> = \text{trace} [\rho ~O]
$$  
In 1877, Boltzmann noticed\cite{Lindley} that he could define a statistical version of the entropy that would agree with both of the other definitions for systems in equilibrium.    This definition gives the entropy as minus the average logarithm of the probability density or density matrix
\be \la{StatisticalEntropy}
S/k_B = < \ln (1/\rho)> =   \text{trace} [\rho ~\ln (1/\rho)]
\ee
Here $k_B$ is the Boltzmann constant, set equal to unity in our other formulas.
The {\em statistical entropy} of \eq{StatisticalEntropy} can be considered to be yet a third kind of entropy.  

Each of these different kinds of entropy has its own sharply defined range for validity: thermodynamic entropy, for large systems in equilibrium,  kinetic entropy for low density gases, that is the  Grad\cite{Grad} limit, statistical entropy, within the range of validity of statistical mechanics.  However each one of these can be extended beyond this narrow range. Both the original ranges and also the extensions overlap.  In fact, the ranges are continually being extended by scientific workers so that the concept of entropy is a growing rather than a static one.  

\renewcommand{\theequation}{3-\arabic{equation}}
\setcounter{equation}{1}

\section{Landau's Kinetic Equation: Boltzmann Reapplied} \la{Landau}
The behavior of excitations in low temperature systems can be described in terms of quasiparticle kinetic equations that are closely analogous to the BKE.  But unlike the BKE, which was mostly invented all at once, the understanding of quasiparticle dynamics was built up step by step over the decades between 1920 and 1960. This chapter traces a little of this development and then concludes with a discussion of entropy within the most transparent of these quasiparticle theories, the Landau theory of the Fermi liquid and its attendant Landau Kinetic Equation (LKE).  
\subsection{Metal physics}
For decades following Boltzmann's studies, the BKE seemed to describe a low density gas, but little else. The requirement that the kinetic equation describe accurately all the collisions that occur in the system could not easily be applied to any of the solids and liquids that surround humankind since al of them clearly involve the simultaneous interaction of many molecules.  However,  the charged excitations moving around in metals did not appear to behave totally differently from gases.  Paul Drude's\cite{Drude1, Drude2} work at the beginning of the Twentieth Century treated the electrical properties of metals in terms that were essentially similar to the kinetic theory ideas of Maxwell and Boltzmann.  The major difference was that the particles in motion in metals were charged and therefore could respond to electromagnetic forces. Looking forward from Drude's day, we can see how these forces might be included by inserting electromagnetic potentials in the left hand side of the Boltzmann equation.  These forces would act continually upon the particles.  The free motion of these particles would be interrupted by occasional collisions so that, as in a gas, the particles would diffuse through the material.   To understand collisions, we might look to modifications of the right hand side of the Boltzmann equation. 

Given the fact that a metal contains a dense periodic lattice of charged ions, each ion capable of strong scattering of electrons, the fact that the scattering would only happen occasionally was, after a time, recognized as quite unexpected.  

One part of the puzzle was solved by Felix Bloch in 1927\cite{Maze}[pages 106-113].  He found that in quantum theory a particle moving in a periodic potential would indeed behave as a free particle\cite{Bloch}.  As in free space, these particles could have  stationary states, each with a well defined energy. Each state would be  labeled by a quantity, $\p$, that behaved much like a  momentum.  In this situation, the energy-momentum relation would no longer have the free space form, $\e=p^2/(2m)$, but instead $\e$ would form a periodic structure with the same periodicities as the ionic lattice of the metal.  So one of the important activities in the physics of the middle of the Twentieth Century was to investigate, understand, and predict the form of this energy function, $\e(\p,\R)$.

The usual metals support very strong Coulomb forces, that can produce bound states with a very large range of electronic energy eigenvalues.  In the usual situation, the temperature is quite small in comparison to this eigenvalue-range.   Early on, it was not clear how the electrons would arrange themselves in this set of configurtions.  
An entirely new insight was put together by many founders of quantum theory including Dirac, Pauli, and Summerfeld  \cite{Maze}[pages 94-97].  They developed the view that electrons were particles that obeyed {\em Fermi-Dirac} statistics and that  for such particles, it is impossible to have more than one particle per quantum state.  So the states tended to fill up, starting from the   lowest levels, which were completely filled,  than a narrow band of partially filled intermediate levels, and then  with the highest levels completely empty.       This situation is not much disturbed by any perturbation. Hence these low-lying  state are inert.  The higher energy state are too high to have an appreciable occupation.  They too remain inert.  It is only the states within a narrow band of energies around a special energy,  called the {\em Fermi energy} that are active.  This band of energies is then the object of study in the electronic part of metal physics.    As in low-density gases, one may describe the fermions in this situation by describing the density of particles labeled by momentum $\p$ in the neighborhood of the spatial point $\R$.  One might as well use the same notation as for a classical gas and describe this density by $f(\p,\R,T)$. 

The major experimental tool for the investigation of the electronic structure involved the placement of the metal in electric and magnetic fields. The observation of the  the resulting motion of the charged carriers could then give considerable insight into the properties of the electronic excitations.   The experimental electric and magnetic fields were almost always slowly varying in space on spatial scales comparable to the distance between ions. They were also slowly varying over the characteristic times  necessary for the quasiparticles to traverse the distance between ions.  Under these circumstances, at low temperatures, the quantum theory for the excitation motion  is is surprisingly simple:   The carriers behave like classical free particles moving under the control of a Hamiltonian.   The Hamiltonian itself is the $\e(\p,\R,T)$ as it is modified by the space- and time- dependent electromagnetic fields in the metal. These fields are produced by both the ions and electrons within the materials and also by slowly varying electromagnetic fields external to the material.  These fields, or rather the  potentials that produce them are reflected in changes in the quasi-particle energy $\e$.  In this energy, the momentum variable $\p$ is enhanced by the addition of electronic charge times the vector potential while the Hamiltonian itself has an additional term of charge times the scalar potential.

\subsection{Quasiparticle excitations} 

 The charge carriers in a metal are correctly described by  this quantum analog of Boltzmann's distribution function, $f(\p,\R,T)$. In the absence of collisions, once the effective Hamiltonian, $\e(\p,\R,T)$, has been constructed, this particles obey Hamiltonian equations of motion in the form as given by Poisson brackets with $\e(\p,\R,T)$.   Thus the entire analysis that gives the left-hand side of Boltzmann's kinetic equation, \eq{BKE}, applies equally to the dilute gas and to metals. Empirically, it is noticed that the charge carriers scatter rather seldom in pure metals at low temperatures. So the collision term is not of primary importance in determining the motion of the electrons.  Excitations that move under the influence of an effective Hamiltonian like $\e(\p,\R,T)$ and which scatter rather infrequently   are termed {\em quasiparticles}.\footnote{The actual word, {\em quasiparticle}, only comes into physics much after the initial work of Bloch.  I have found the word in a 1947 paper of N. N. Bogoliubov on the behavior of dilute gases composed of Bose  particles.\cite{Bogoliubov}} Since quasiparticles provide a correct description of the low temperature behavior of many materials, quasiparticle analysis permeates condensed matter physics.

\subsection{Quasiparticle scattering: The right hand side}

Through the 1920s and beyond, scattering rates like the one Maxwell and Boltzmann used were employed to understand the scattering of particles, quasiparticles, and other excitations.  The relevant quantum mechanical formula came under the name of the ``Fermi's golden  rule.'' It said that whenever the potential causing the scattering was weak the scattering rate could be computed\footnote{Much of the ideas for this comes from the work of Dirac\cite{DiracGolden}.} by
\bea \la{GoldenRule}
&~&\text{rate}= \sum_\text{different possibilities} \frac{2 \pi}{\hbar} ~\times~  \text{delta function for energy conservation}
\nonumber \\
 &\times& \text{~density of initial states~} \times~|<i|~V~|f>|^2 ~\times \text{~density of final states~} 
 \nonumber \\ &~&
\eea
Here $<i|~V~|f>$ is the matrix element between initial and final states of the potential causing the scattering. Since the squared matrix element typically contains a delta function for momentum conservation we can recognize this golden rule expression as one possible source of the $\mathcal{P}$ we  used in the Boltzmann kinetic equation. 

Since the early days of quantum mechanics, this approach has been 
used to understand the behavior of a continually expanding range of systems, including those of condensed matter.  As this approach was developed, it was recognized that the scattering of the quasiparticles would end up as being quite important. The scattering might be infrequent, but they were necessary to bring the system into thermodynamic equilibrium.  

In dilute ordinary gases,  Boltzmann's basic scattering approach could be carried forward to work within quantum theory.     The scattering rate would be calculated by quantum theory, using a formulation like \eq{GoldenRule} with differential cross-sections replacing the squared matrix element. 
Thus,  the basic structure of the kinetic calculation remained the same as the one introduced by Boltzmann. 

\subsubsection{Degenerate scattering rate}
The most important addition that quantum theory brought to BKE type calculations came with the recognition that scattering processes would be modified to include ``quantum statistics''.  Quantum mechanical particles come in two kinds:  fermions, which do not admit of more than one particle existing in a single quantum mode and bosons, in which the occupation of modes by multiple particles (or particle-like excitations) is enhanced.    The very complex behavior which can be produced by these statistics nonetheless is reflected by very simple changes in the collision terms used in Boltzmann-like analysis.  In a translationally invariant system, one can, as before say that the modes of the system can be labeled by momentum vectors, and that the number of modes available per unit volume in a volume $d\p$ in momentum space is $d \p/(2\pi \hbar)^D$, where $D$ is the spatial dimension  and $\hbar$ is Planck's constant.  We shall forget about the denominator, which can be absorbed into the normalization of the scattering rates $\mathcal{P}$ and $\mathcal Q$.  So once more, we shall use the momentum vector $\p$ to describe the scattering processes.  In our further description we shall use, as before, $f(\p,\R,T)$ to describe the density of particles, but now our normalization will make $f$ the average number of particles in each of the quantum modes in the neighborhood of the momentum $\p$.     For Fermi-Dirac particles the density of available states will be reduced by the factor $1-f(\p,\R,T)$, to reflect the fact that multiple occupation is impossible.  Bose-Einstein\cite{Bose,Einstein,Einstein1} particles have their scattering into a given range of modes enhanced by a corresponding factor of $1 + f(\p,\R,T)$.   To cover both cases, and to include nondegenerate situations, we shall take the scattering rate to be modified by the factor 
\be
f^>(\p,\R,T)=1 + \varsigma f(\p,\R,T)  
\ee
with  $\varsigma =1,0,-1$ respectively for bosons, nondegenerate particles, and fermions.    The $f^>$ so defined is the density of states for the excitations involved.  In the case of fermions, it can also be described as the density of holes. For bosons this factor describes the enhancement of scattering which, for radiation, is described by the words {\em stimulated emission}\cite{Einstein}. For symmetry, sometimes when we describe the density of excitations, $f$, we shall write it as $f^<$.
\subsubsection{Variational approach}
Classical formulas for scattering rates were defined in \se{ClassicalVariational}. 
A direct extension of our variational derivative description of BKE scattering as given in \eq{ClassicalScattering}  is to define a $\psi$ to be a functional of both the density of quasi-particles and also the density of states for these quasiparticles, $f^>=1+\varsigma f,$.  In this writing we have
\bsubs \la{varyQuantum}
\bea\la{psiQuantum}
\psi[f^<,f^>] &= & \frac 12 \int  ~ d\p~d\p'~d\q~d\q'~ \nonumber \\
&~& f^<(\p) f^<(\q)  f^>(\p')  f^>(\q') \nonumber \\
&~& \mathcal{Q}\left(\begin{array} {ccc} \p &  \rightarrow& \p' \\
\q &  \rightarrow& \q' \\
 \end{array}\right)  \nonumber \\
 &~&\d(\p+\q-\p'-\q') \d\big(\e(\p)+\e(\q)-\e(\p')-\e(\q')\big)
\eea
and then get scattering rates via formulas analogous to \eq{s>} and \eq{s<}
\bea  \la{s>p}
\s^>(\p,\R, T)  &= & 
 \left. 
 \frac {\d \psi[f^<,f^>]}{\d f^<(\p)}
\right|_{f^<(\cdot)=f(\cdot,\R,T),f^>(\cdot)=1+\varsigma f(\cdot,\R,T)}
\nonumber \\ 
&=&  \int ~d\q~   d\p'~d\q'  ~\mathcal{Q}\left(\begin{array} {ccc} \p &  \rightarrow& \p' \\
\q &  \rightarrow& \q' \\
 \end{array}\right) \nonumber \\ 
&~&  ~ f(\q,\R,t)~[ 1+ \varsigma f(\p',\R,T)] [ 1+ \varsigma f(\q',\R,T)]
\nonumber \\
 &~&\d(\p+\q-\p'-\q')  \nonumber \\ &~&
\d\big(\e(\p,\R,T)+\e(\q,\R,T)-\e(\p',\R,T)-\e(\q',\R,T)\big) \nonumber \\
\eea 
as well as 
 \bea  \la{s<p}
\s^<(\p',\R, T)  &= &  \left. 
 \frac {\d \psi[f^<,f^>]}{\d f^>(\p')}
\right|_{f^<(\cdot)=f(\cdot,\R,T),f^>(\cdot)=1+\varsigma f(\cdot,\R,T)}
\nonumber \\
&=&  \int d\p ~d\q~  d\q'  ~
\mathcal{Q}\left(\begin{array} {ccc} \p &  \rightarrow& \p' \\
\q &  \rightarrow& \q'  \nonumber \\
 \end{array}\right)     \nonumber \\
&~& f(\p,\R,T) ~ f(\q,\R,t)~  [ 1+ \varsigma f(\q',\R,T)] 
\nonumber  \\
 &~&\d(\p+\q-\p'-\q')   \nonumber \\ &~&
  \d\big(\e(\p,\R,T)+\e(\q,\R,T)-\e(\p',\R,T)-\e(\q',\R,T)\big) \nonumber \\
 \eea 
 These results hold in second order perturbation theory with a two-body potential 
 $$V(1,2)=\delta(t_1-t_2) \int ~ d\p ~\exp[i\p\cdot (\mathbf{r}_1-\mathbf{r}_2)] ~\tilde v(\p)$$
 Then $\mathcal{Q}$ will have the form
 \bea  \la{second}
 \mathcal{Q}\left(\begin{array} {ccc} \p &  \rightarrow& \p' \\
\q &  \rightarrow& \q'  \nonumber \\
 \end{array}\right)  \sim \tilde v(\p-\p')\big[\tilde v(\p'-\p)+\varsigma \tilde v(\q'-\p) \big]
 \nonumber \\
 \eea
\esubs       
Note that this result is automatically even (or odd) under the interchange of $\p'$ and $\q'$ for bosons (fermions).  

\subsubsection{Are quasi-particles well-defined?}
Excitations in a translationally invariant non-interacting system may be classified by their momentum $\p$.  With, for example, periodic boundary conditions this momentum is quantized so that one can keep track of individual single-particle states.  Bloch showed that this is true also if the system has a static periodic potential.  The idea of quasiparticles is based upon the view that as we gradually turn on interactions, these single-particle states will deform, but they will not change in any essential way. We know that this gentle deformation does not always hold.  For example, if we break the periodicity by adding static impurities to a metal, it can contain localized single-particle states, with properties entirely different from the non-interacting excitations.  Nonetheless, in the absence of evidence to the contrary, physicists usually assume that a translationally invariant material can be described by excitations with a well-defined momentum.

It turns out that at moderately low temperatures, this quasiparticle description is roughly true for a wide variety of materials.  Its validity can be checked in part by calculating or estimating the inverse lifetime of the excitation using the scattering formulas of \eqs{varyQuantum}.  If this width times $\hbar$ is much smaller than the range over which one has significant variations in the  quasiparticle energy, then the quasiparticle point of view is likely to work.  For example in a Fermi  liquid (in particular He$^3$) the width of the active band of excitations is proportional to temperature, but the scattering broadens each excitation by only an amount proportional to temperature squared.  Therefore, at lower temperatures, a quasiparticle theory is likely to be valid.   

At the very lowest temperatures however all kinds of new dynamical phenomena are possible, so present theory has nothing definitive to say. 

\subsection{Quasi-particle energy: The left hand side}
In 1956 Landau extended the ideas we are discussing by introducing a theory of the Fermi  liquid, in part for application to the behavior of He$^3$ at low temperatures\cite{L3,L4, L5}. The first of these papers is the one that defines the Landau Kinetic Equation and  will be the paper relevant for our present discussions. The remarkable and new element of the LKE was the determination of the quasi-particle energy  from the thermodynamic free energy.
According to Landau, who offered us the equations to describe the Fermi  liquid but did not say how they might be justified, the free energy, $\mathcal{G}$, is a functional of the distribution function $f$ such  that the quasiparticle energy may be calculated as
\be
\e(\p,\R,T)= \frac{\d \mathcal{G}}{\d f(\p,\R,T)}.
\ee 
The quasiparticle energy, thus defined, can then be used within the Poisson bracket in a Boltzmann equation, i.e. \eq{LKE}, conceptually and structurally identical to the original BKE of \eq{BKE}.  
\bea
\la{LKE}
\frac{\partial f(\p,\R,T)}{\partial t}&+ &\{\e(\p,\R,T), f(\p,\R,T)\}
\nonumber \\
&=& - \s^>(\p,\R,T)f(\p,\R,T) +\s^<(\p,\R,T) [1+\varsigma f(\p,\R,T)]
\nonumber \\
\eea
\eq{LKE} is the Landau Kinetic Equation, LKE. 

The predictions from Landau's theory and particularly the LKE were soon bourn out by the experimental data\cite{Wheatley}.    
\subsubsection{ ``Golden Rule'' for correlation energy}
Landau's free energy includes the effects of correlations produced by the interactions among the particles. In parallel to the construction of a golden rule for interparticle scattering, there is an analogous and equally useful golden rule for correlation energy.  Once again this result applies most directly to the case of relatively weak interactions among particles, but can be extended to apply, at least in a qualitative fashion, much more generally.   In the golden rule calculation, which is essentially a result of second order perturbation theory,
the correlation energy is of the form
\bea \la{phiQuantum}
\phi[f]&=&\frac 1\pi ~\text{Pr}
\int ~d\p ~d\q ~d\p'~ d\q' ~ \mathcal{Q}\left(\begin{array} {ccc} \p &  \rightarrow& \p' \\
\q &  \rightarrow& \q'  \nonumber \\
 \end{array}\right)  \nonumber \\
&~&  f(\p) f(\q) [1+\varsigma f(\p')] [1+\varsigma f(\q')] 
 \nonumber \\
&~& \frac{ \delta(\p+\q-\p'-\q')}{\e(\p) +\e(\q)-\e(\p') -\e(\q')}
\eea

The only difference from the generator of the collision terms as seen in \eq{psiQuantum} is that \eq{phiQuantum} contains an energy denominator in place of \eq{psiQuantum}'s energy delta function.  The variational derivative of the expression of \eq{phiQuantum} with respect to $f$ gives the correlation contribution to the single-particle energy $f$. 
\subsection{Quasiparticle entropy}
The calculation of the time-dependence of the entropy density from Landau's version of the kinetic equation, \eq{LKE} is most simple and instructive. Although Landau makes use of the equation for the entropy derived below, for some reason he assumes its truth {\em a priori} and does not derive it from his kinetic equation.  Let us imagine that we wish to calculate the time dependence of any function of the distribution function say $\mathcal S(f)$, which has a derivative\footnote{Here ``S'' is the standard symbol for entropy while ``M'' stand for multiplier.}   
\be \la{der}
\mathcal{M }(f) = \frac {d\mathcal S(f)}{df } 
\ee
Multiply \eq{LKE} by $\mathcal{M }(f(\p,\R,T) $ and integrate over all values of $\p$.   The result is exactly of the form of our previous equation, \eq{DefineEntropy}, for the entropy density
\bsubs
\be \la{QuasiparticleEquation} 
\partial_T~ \rho_s(\R,T)  +\nabla_\R \cdot j_s(\R,T) =  \text{RHT}_s 
\ee
The equation describes what we now term as an  entropy density
\be  \la{QuasiparticleEntropyDensity} 
\rho_s(\R,T)   =   \int d\p ~ \mathcal S(f(\p,\R,T))  
\ee
and an entropy current density
\be \la{QuasiparticleCurrentDensity} 
 j_s(\R,T)  = \int d\p ~ \mathcal S(f(\p,\R,T)) ~ (\nabla_\p \e(\p) ) 
\ee
\esubs
So far we need not have specified the functions $\mathcal{M }$ and $\mathcal{S }$. \eq{QuasiparticleEquation}  is a correct statement for any value of the function $\mathcal S(f)$.The defining  statement for the entropy is that the collision term for the observable, here described as the right hand term RHT$_s$, must always be greater than zero, except in local equilibrium where it will be zero. 
There is a choice that will meet the requirement that the RHT$_s$  be non-negative in this fashion.  That choice, unique up to an additive constant, is
\bsubs
\be
\mathcal{M }(f)=-\ln \frac{f}{1+\varsigma f} 
\ee 
which then  implies
\be  \la{S}
 \mathcal{S}(f)  =-f \ln{f} +\varsigma (1+\varsigma f) \ln{(1+\varsigma f) }
\ee
\esubs
This choice will make the collision term have the form
\bea \la{RHT_S}
RHT_s &=& \frac 12\int ~d\p~d\q~   d\p'~d\q' 
~ \mathcal{P}\left(\begin{array} {ccc} \p &  \rightarrow& \p' \\
\q &  \rightarrow& \q' \\
 \end{array}\right)  
 \nonumber \\
 &~&   \ln \frac { f(\p) ~ f(\q) ~[1+\varsigma f(\p')] [1+\varsigma f(\q')]}{[1+\varsigma f(\p)] [1+\varsigma f(\q)]f(\p') ~ f(\q')}
 \nonumber \\ 
 &~& f(\p)  f(\q) [1+\varsigma f(\p')] [1+\varsigma f(\q')] 
\eea
which will then force this term to be non-negative whenever $\mathcal{P}$ obeys the detailed balance symmetry of \eq{DetailedBalance}.  (See the discussion after \eq{RHT_sc}.)  The only possibility of a zero is when
$ f(\p) /[1+\varsigma f(\p)] $ is proportional to an exponential with exponent being a sum of local coefficient times the conserved quantities $O=1, \p$ and $ \e$.

As has long been known, the quasiparticle result binds together the different definitions of entropy.   The calculation is a kinetic one, having the property that it holds into situations well out of equilibrium. It appears to remain true in all situations in which the quasiparticles are well defined.   Of course the quantity defined by \eq{QuasiparticleEntropyDensity} and \eq{S} is also a thermodynamic entropy since it works for equilibrium situations.  Further,  quasiparticle theory agrees with  with the various laws of thermodynamics and part of that agreement is that the spatial integral of the quasiparticle entropy density, as present in \eq{QuasiparticleEntropyDensity}  serves as the thermodynamic entropy within that thermodynamics.   Finally, this entropy fully agrees with a count of the number of quasiparticle configurations, as presented for example in Schr\"odinger's  book on statistical physics\cite{Schrodinger}. Everything fits together under one roof.

\renewcommand{\theequation}{4-\arabic{equation}}
\setcounter{equation}{0}

\section{ Seeking Quantum Kinetic Entropy} \la{Green}
\subsection{Introducing thermodynamic Green's functions}
\subsubsection{To a one-body analysis}
The general form of a quantum non-equilibrium analysis is very complex indeed.  For $N$ spinless particles one must specify a density matrix that is a function of $6 N$ separate variables when our system exists  in the usual three dimensional space.  However, analysis suggests that simpler situations are possible. In the last chapter we saw a quasiparticle state is specified by $f$, a function of only six variables.  At full equilibrium at a given temperature a  state-specification requires of order a half-dozen thermodynamic parameters. 

One possible theory for describing both equilibrium and non-equilibrium  situations uses Green's functions, $G(1,2)$. These functions are expectation values of creation and annihilation operators.    This original equilibrium Green's function theory specified by Martin and Schwinger \cite{MS} is based 
upon a time boundary condition due to Kubo\cite{Kubo}.    The result is that the Green's 
function, $G$, is composed of two 
functions, $G^<$ and $G^>$, that both have a time dependence in which they depend only upon only upon the time difference variable, $t_1-t_2$, with the times sitting in an interval $[0,-i\b]$.   The function,  G(1,2), can be split into two functions, $G^<$ and $G^>$, with times that can be analytically continued onto the real line.   These functions can be defined by a formally exact diagrammatic analysis  using a Dyson equation\cite{Dyson,Greens} and a set of skeleton diagrams that depict the Green's function and the two-body potential.   This equilibrium analysis employs one- and two-body potentials that are both functions of two spatial variables in addition to the time difference..  Once these potentials are given, the thermodynamic state is defined by  roughly a half dozen equilibrium parameters, including temperature and chemical potential. Possible behaviors include many thermodynamic phases as well as various kinds of band structures and localized states.  Many of these situations would give a behavior very different from that described by the Fermi liquid theory. 

This Martin-Schwinger theory may be extended to describe time dependent phenomena, via a generalization of  these same Green's functions.  Followup theory (See \cite{Schwinger, KB, Keldysh}.) suggests that if one starts at equilibrium at some initial time, in a system specified by one and two particle potentials, $U(1,2)$ and $V(1,2)$, then for subsequent times the system can still be be fully described by a one particle Green's functions. These functions are specified by an equation of motion and further by a perturbation expansion containing an infinite set of diagrams.   This situation is once again described by two Green's functions, now written as $g^<(1,2)$ and $g^>(1,2)$, that  depend upon both time variables, $t_1$ and $t_2$. A perturbation theory analysis once more generates a formally exact description of a self-energy to define these Green's functions.  I believe that time dependent situations are potentially more complex than static ones, but  I do not know the degree to which this non-equilibrium situation can display a richer complexity of different phases or behaviors than is shown by the equilibrium case.

We are looking for a Boltzmann-like analysis. An additional condition is required for this approach to to apply.  The functions $g^>, g^<, U$ and $V$ are all functions of a pair of space-time variables $(1,2)=(\r_1 t_1,\r_2 t_2)$. These can then be expressed in terms of sum and difference variables as
\be \la{variables}
\R=\frac 12( \r_1+\r_{2} ), T=\frac 12(t_1+t_{2})    \text{~~and~~}   \mathbf{r}  =  \r_1-\r_{2},t= t_1-t_{2}
\ee
As in the usual analysis of Wigner functions\cite{Wigner}, a quasi-classical situation arises whenever the variation in the sum variables is much slower than that in the corresponding difference variables. In physical terms, the spatial variation of the sum variable must be slow in comparison to typical microscopic distances, including force ranges and than de Broglie wave lengths. The time variation in the sum variable must also be sufficiently slow so that we can consider all the excited energy levels in the system to be part of continua, rather than discrete levels.   In this quasi-classical situation, the Green's functions obey an equation very analogous to Boltzmann's, with the important difference that excitations are described by both a momentum, $\p$, and a frequency, $\o$.  Since there is a close analogy between the behavior of these two variables we shall often group them together as an energy-momentum variable, $p$.

\subsubsection{To a generalized Boltzmann equation}
An excellent summary of the subject of this section is provided by Joseph Maciejko\cite{Maciejko}. For a summary of historical development and of applications see \cite{Green's_Functions} especially the articles by Baym and Martin.

As pointed out in ref.\cite{Maciejko}[page 21] to derive the basic equation of the quantum kinetic theory one begins by assuming that the two forms of the Dyson equation:
\be \la{Dyson}
[G_O^{-1}-\S]~G =1  \text{~~and~~} G ~[G_O^{-1}-\S] =1
\ee
both give the same (correct) answer for $G$ when they are applied in the time interval $[0,-i\b]$.  

To do real time analysis of these equations, one uses Wigner functions\cite{Wigner}. These are time-ordered versions of $G$ with times on the real axis. They are expressed in sum and difference variables, as in \eq{variables}, and then Fourier transformed with respect to the different variables. The resulting real  functions are  $g^<(\p,\o,\R,T)$, which  describes the density of excitations labeled by $\p,\o$ in the space-time neighborhood of $\R,T$,  and $g^>(\p,\o,\R,T)$, which gives the corresponding density of available states in these variables. For fermions, $g^>$ can also be considered the density of holes.  A density of states, uncorrected by occupancy,  is given by
\bsubs
\be
a(\p,\o,\R,T) = g^>(\p,\o,R,T)- \varsigma~ g^<(\p,\o,R,T)
\ee
(Recall that $\varsigma$ is plus one, minus one or zero respectively for bosons, fermions and particles obeying classical statistics.) For bosons, $g^>$ is greater than $a$, indicating the enhancement of scattering caused by state occupation. For fermions, $g^>$ is smaller than $a$. Finally, we shall make use of 
the propagator
\be
g(\p,\o,\R,T) = \text{Pr} \int \frac{d\o'}{2\pi} \frac{a(\p,\o',\R,T) }{\o-\o'}
\ee
\esubs
with  Pr indicating a principle value integral. 

The Dyson self-energy $\S$ has a formally similar description. We define $\s^<(\p,\o,R,T)$ and $\s^>(\p,\o,R,T)$ as the two real-time, time-ordered components of $\s(1,1')$.  We use subsidiary quantities
\bsubs
\be
\gamma(\p,\o,\R,T) =\s^>(\p,\o,R,T)- \varsigma \s^<(\p,\o,R,T)
\ee
and the ``real'' (rather than complex) self-energy
\be
\s(\p,\o,\R,T) = 
\s_{HF}(\p,\R,T) +\text{Pr}\int \frac{\d\o'}{2\pi} \frac{\gamma(\p,\o,\R,T) }{\o-\o'}
\ee
\esubs
Here, $\s^>(\p,\o,\R,T) $ is defined as the rate of scattering of an excitation out of a configuration  $\p,\o$ in the neighborhood of $\R,T$, while $\s^< $ is, after being multiplied by the density of states,  the corresponding scattering rate into that configuration.  Whenever the two-body potential, $V$, is a delta function in time, with space Fourier transform, $\tilde{v}$,    the {\em Hartree-Foch} contribution to the self-energy,  denoted as $\s_{HF}$, is independent of frequency and cannot be represented in the same fashion as the rest of the self-energy. It  has the form \cite{KB}[pages 17-26]  
\bea \la{HF}   
\s_{HF}(\p, \R,T)= \int  d\p'~ d\o' [\tilde{v}(0) +\varsigma \tilde{v}(\p-\p')]g^<(\p',\o',\R',T)
\eea
in equilibrium in a translationally invariant situation. We shall henceforth assume this frequency-independent form for the Hartree-Foch term.  The independence will cause it to disappear from many of our results.

\subsubsection{Equations of Motion}  
We can see these definitions manifested in the generalized kinetic equations, which are obtained by subtracting the two forms of the Dyson equation, \eq{Dyson} and then doing a Fourier transform in the difference variables.   In a first order gradient expansion, the resulting equations then read\cite[page 27]{Maciejko}   
\bsubs
 \la{KB}
 \be
 \la{g<}
[g_0^{-1}-\s,g^<]-[\s^<,g] = -\s^> g^<+\s^< g^>
\ee
and
\be 
\la{g>} 
[g_0^{-1}-\s,g^>]-[\s^>,g] = \varsigma[-\s^> g^< + \s^< g^>]
\ee
\esubs
with all quantities being functions of $\p,\o,\R,T$ 
and with the generalized Poisson bracket defined as\footnote{Note that the $\p,\R$ part of this bracket has the opposite sign from the one used in the bracket we employed for the BKE and LKE.}  
\be \la{bracket}
[a,b]= \frac{\partial a}{\partial \omega} \frac{\partial b}{\partial T}
      -\frac{\partial a}{\partial \p} \frac{\partial b}{\partial \R}
     -\frac{\partial a}{\partial T} \frac{\partial b}{\partial \o}
     +\frac{\partial a}{\partial \R} \frac{\partial b}{\partial \p}
\ee
\eqs{KB} have corrections of second order in space and time derivatives.  To a similar accuracy, we obtain, by adding the two forms of Dyson's equations, \eq{Dyson}, to obtain\cite[page 27]{Maciejko}

\be   
g = \frac{g_0^{-1}-\s}{D}\text{~~and~~}
a=\frac{ \gamma}{D}
\text{~~with~~} D= (g_0^{-1}-\s)^2+(\gamma/2)^2
\ee
These solutions for $a$ and $g$  are consistent with the result of taking the difference between second form of the kinetic equations \eqs{KB} and $\varsigma$ times the first form, which then gives
\be \la{identity}
\big[{g_0}^{-1}-\s,a\big]-\big[\gamma,g\big]=0
\ee

\eqs{KB} can be interpreted as simple generalizations of Boltzmann's kinetic equation.    On the right we see scattering terms, respectively describing the rates of scattering out of and into a configuration described by $\p,\o,\R,T$. The only thing that is new is the inclusion of the energy label, $\o$ in addition to the momentum label $\p$.    On the left hand side the generalized Poission bracket
$[g_0^{-1}-\s,g^<]  $ describes the rate of change of $g^<$ caused by gradual changes in particle position, momentum, and energy. Here $g_0$ is the propagator for the non-interacting system.  It may include electromagnetic vector and scalar potentials as well as a one body potential $U(1,2)$. The self-energy, $\s$, includes all modifications of single-particle energies from particle interactions.     

 Finally, the equation for $g^<$, \eq{g<}, contains the term $[\s^<,g]$. This term describes the  changes in particle density produced by the space-time variation in the rate of addition of particles to this configuration via $\s^<$.  The change is brought to $g^<$ by  the propagator,  $g$.   This {\em backflow term} in the generalized Boltzmann equation thus describes how information from the collision is brought back to the distribution function, $g^<$. This term, and its analogs in more accurate approximations,\footnote{For example, Ref \cite{Maciejko} describes how spin is included by making \eq{g<} and \eq{g>} into matrix statements, with anti-commutators on the right and commutators on the left.} might be  the source of complexity related to the entanglement of quantum states.  Usually, we believe that entanglement is connected with phases of complex wave functions.   However, because the $g$'s are real these phases cannot appear directly.   Nonetheless a tremendous amount of physical complexity can be encoded in functions of four variables so some functional analog of entanglement might well  be possible.   
 
The equations just presented provide a complete theory of the dynamics of degenerate (or non-degenerate) quantum systems slowly varying\footnote{Notice that the ``functional analog'' of entanglement mentioned in the previous paragraph might well involve the generation of short-range correlations, thereby voiding the conditions required for thre validity of the kinetic equations.} in $\R,T$.  Relatively simple formulas can be used to find many physical quantities.  The simplest example, coming directly from the meaning of the Green's function,  is that the density of particles, momentum, and energy are\cite{KB}[page 127] given by or closely related to the integral over $\p,\o$ of $g^<(\p,\o,\R,T) $ multiplied by, respectively, $1,\p$, and $\o$.  We shall discuss a few of these connection below.  In addition, the Green's functions  yield direct expressions even in  non-equilibrium situations for such quantities  as correlation energy,  stress tensor, and partition function\cite{BaymConserving}. From these we can derive a non-equilibrium counterpart of  temperature times entropy.\footnote{H.C. \"Oetttinger\cite{Oettinger} has also emphasized the necessity for defining temperature in order to pull entropy out of a dynamical argument.   However, we have noticed that Boltzmann's and Landau's theories permit a simple definition of kinetic entropy even when temperature is undefined.}
However, as we shall see, entropy\footnote{Recent  work \cite {Srednicki,Deutsch} describes equilibrium behavior in energy eigenstates without invoking an analysis involving something like an H-theorem}    itself remains elusive as a general concept in the non-equilibrium part of  this Green's function formulation.

\subsection{The Maths}
\subsubsection{Hilbert transforms}
A crucial element of the definitions just given is the Hilbert transform  of a function of time, $t$. Given a function $F(t)$ with Fourier transform $f(\o)$ defined by 
\be 
f(\o) = \int dt \exp(i \o t)~ F(t) 
\ee
Let $\text{sign}(t)$ be the function that is plus one for positive $t$ and minus one for negative $t$.  Now one can see that the Fourier transform of sign$(t)F(t)$ is
\be \la{Hi}
\int dt \exp(i \o t)~ F(t)~ \text{sign}(t)=\text{ Pr} \int \frac{d\o'}{2 \pi} \frac{2i}{\o-\o'}f(\o')
\ee
 In Fourier space, then, the Hilbert transform is  given by a convolution with  $2/(\o-\o' ) $.  In real space it is defined by multiplication by $-i~$sign$(t)$.   Thus, if we apply the Hilbert transform twice we get minus our starting point.  As a result, 
\be  \la{Hilbert2}
\text{Pr}  \int \frac{d\o}{2\pi} \frac{1}{(\o-\o_1)(\o-\o_2) } =\frac{\pi}{2}~ \delta (\o_1-\o_2)   
\ee
We shall use \eq{Hilbert2} in what follows.

\subsubsection{Bracket structure} 
The analysis of the left and side of the Boltzmann equation will require considerable manipulation of the Poisson brackets.  These brackets obey a set of standard identities including
\bsubs
\begin{itemize}
\item{antisymmetry}
\be [A,B]=-[B,A] \ee 
\item{derivative structure}
\be [A,B~C]= [A,B]~C +[A,C]~B \ee
\item{Jacobi identity}
\be [[A,B],C] +[ [B,C],A]+[[C,A],B]=0  \ee
\end{itemize}
\esubs
In addition, we shall need an identity which applies under integrations with respect to $\p$ and $\o$, namely 
 the {\em bracket identity}
\bea \la{BracketIdentity}
&&\int ~ dp ~\big\{~ [A ,B] C-A[B,C]~\big\} = \nonumber \\
&&~~~~~~~~
 - \partial_T  \int ~ dp ~ A ( \partial_\o B) C  
+\nabla_\R \cdot \int ~ dp ~A ( \nabla_\p B ) C
\eea
We shall describe the situation, as on the left of \eq{BracketIdentity}, in which an integral over p  results in an expression in the form of a perfect derivative, 
$$
\partial_T\rho+ \nabla_\R \cdot j 
$$  
as a {\em perfect integral}.  As indicated in \se{alternative}, this perfection  is important in the derivation of conservation laws. By setting $C=1$ in \eq{BracketIdentity}, we notice that the $p$ integral of a bracket is always a perfect integral
\be \la{bracket_identity}
\int ~ dp ~ [A ,B]= - \partial_T  \int ~ dp ~ A ( \partial_\o B)  
+\nabla_\R \cdot \int ~ dp ~A ( \nabla_\p B )
\ee

\subsubsection{ $\Phi$ derivability}

 The generalized Boltzmann equation is algebraically complex, but its analysis can be considerably simplified  by using Green's function approximations that are {\em conserving}\cite{BaymKadanoff} and {\em $\Phi$-derivable}\cite{BaymConserving}.
 
An important advance in the analysis of Green's function approximations came  from the work of Baym\cite{BaymConserving} who built upon earlier work on conserving approximations for Green's functions\cite{BaymKadanoff, Kraichnan} by applying  Luttinger-Ward\cite{LuttingerWard} studies of thermodynamic consistency within Green's function schemes.   This last built upon a specific scheme for doing perturbation analysis  via a self-energy  that would be expanded in terms of $G$'s and $V$ in the form of {\em skeleton diagrams}.  The particular expansion used started with a functional of $G$ and $V$ defined by Luttinger and Ward\cite{LuttingerWard}, called  $\Phi[G]$ that would generate the self-energy via a variational derivative in the form 
\be \la{Sigma}
\S(1,2)= \frac{\d\Phi}{\d G(1,2)} \ee
Gordon Baym and I \cite{BaymKadanoff} showed that Green's function approximation schemes using variational derivative methods automatically agreed with the standard conservation laws for particle number, energy, and momentum\cite{KB,BaymKadanoff}.  Further, 
Baym\cite{BaymConserving} then built an analysis of Green's function's in the time interval $[0,-i~\b]$ using the generator of \eq{Sigma} to guarantee that the Green's-functions  evaluations of  thermodynamic properties would agree with thermodynamic identities.   Thus these approximations would then give a consistent representation of hydrodynamic behavior.   

A large group of applications have arisen from this thread of analysis\cite{Green's_Functions}, particularly from the generalized Boltzmann equation approach.   For example, one consequence of this line of argument is that the correlation energy  is given by 

\be \la{Correlation Energy}
 2 E_{corr} = \int d1~d2 ~ G(1,2) ~ \frac {\delta \Phi }{\delta G(1,2)}
\ee 
This result arises directly from the equation of motion for $G$. 

\subsubsection{The simplest generating function}

We shall show how this all works by using a very simple generating function for deriving the $\s$'s. The direct extension of our previous generators collision rates of \eq{varyQuantum} and \eq{ClassicalScattering} are the expressions:
\bsubs  \la{simplest}
\bea     \la{simplest<}
\s^<(p) &=&  \int {dq~dp'~dq'}~g^<(q)~g^>(p')~g^>(q') 
\nonumber \\
&& \mathcal{Q}
\left(\begin{array} {ccc} p &  \rightarrow& p' \\
q &  \rightarrow& q' \\
 \end{array}\right) 
 \delta(p+q-p'-q') 
\eea
\bea  \la{simplest>}
\s^<(p') &=&  \int {dp~dq~dq'} g^<(p)~g^<(q)~g^>(q') 
\nonumber \\
&& \mathcal{Q}
\left(\begin{array} {ccc} p &  \rightarrow& p' \\
q &  \rightarrow& q' \\
 \end{array}\right) 
 \delta(p+q-p'-q') 
\eea

In direct analogy with our previous work we note that these collision rates can be directly determined by variational derivatives 
  \be
\s^<(\p,\o') =\frac{\d\psi[g^>,g^<]}{\d g^>(\p,\o)} \text{~~and~~} \s^>(\p,\o') =\frac{\d\psi[g^>,g^<]}{\d g^<(\p,\o)} 
\ee
with the variational function 
\bea \la{simplestpsi}
\psi[g^>,g^<]  &=& \frac1{2 } ~\int dp~dp'~dq~dq'   g^<(p)~g^<(q)~g^>(p')~ g^>(q')  \nonumber \\
&& \mathcal{Q}
\left(\begin{array} {ccc} p &  \rightarrow& p' \\
q &  \rightarrow& q' \\
 \end{array}\right) 
 \delta(\p+\q-\p'-\q') 
\eea
\esubs

This choice is fine for the analysis of the right hand side of the Boltzmann equation. However, the left hand side of our generalized Boltzmann equation depends upon the Hilbert transform of the scattering rates. A slightly different generator will work better for the analysis of this side. This choice probably can be better generalized to higher orders of skeleton graph perturbation theory than the $\psi$ of \eq{simplest}.  For completeness, we include the Hartree-Foch contribution (See \eq{HF}.) to the self energy. We therefore define a variational function that is a sum of Hartree-Foch and correlation terms
 \bsubs  \la{simplestphi} 
\be
\phi=\phi_{HF}+\phi_c 
\ee
with the Hartree-Foch term
\be \la{phiHF}
\phi_{HF}[g^<]=\frac 12  \int ~ dp ~dq ~g^<(p) g^<(q)[\tilde{v}(0)~ 
+\varsigma ~\tilde{v}(\p-\q)] \ee
and the correlation term 
\bea\phi_c[g^>,g^<]  &=& \frac1{ \pi} ~\text{Pr}~\int dp~dp'~dq~dq'    \nonumber  \\
&~&\frac{ g^<(p)~g^<(q)~g^>(p')~ g^>(q')} {p_0+q_0-q_0'-p_0'}   \nonumber \\
&& \mathcal{Q}
\left(\begin{array} {ccc} p &  \rightarrow& p' \\
q &  \rightarrow& q' \\
 \end{array}\right) 
 \delta(\p+\q-\p'-\q') 
\eea
\esubs
since we would then have a situation in which there is a frequency denominator in the correlation term.  The Hartree-Foch term has to be handled specially since it is frequency independent while the Hilbert transform of a frequency independent function is ill-defined.  
Using this structure of \eq{simplestphi}, we can --as will be seen-- avoid most of the complication of working with the LHT.  Using \eq{simplest} and \eq{Hilbert2}, we find that the previously defined scattering rates of \eq{simplest>} and \eq{simplest<} equally come from this variational function
\bsubs   
\be 
\s^<(p) =\text{Pr} \int \frac{d\o'}{2 \pi}~ \frac{1}{\o-\o'} ~  \frac{\d\phi[g^>,g^<] }{\d g^>(\p,\o')}    
\ee
and
\be
\s^>(p) =-\text{Pr} \int \frac{d\o'}{2 \pi}~ \frac{1}{\o-\o'} ~  \frac{\d\phi_c[g^>,g^<] }{\d g^<(\p,\o')}    
 \ee
From \eq{Hilbert2} it immediately follows that 
\be  \la{t<} 
\frac 14 ~ \frac{\d\phi[g^>,g^<] }{\d g^>(\p,\o)} =-\text{Pr} \int \frac{d\o'}{2 \pi}~ \frac{1}{\o-\o'} ~ \s^<(\p,\o')  =t^<(p) 
\ee
and
\be \la{t>}
\frac 14 ~\frac{\d\phi_c[g^>,g^<] }{\d g^<(\p,\o)}  =\text{Pr} \int \frac{d\o'}{2 \pi}~ \frac{1}{\o-\o'} ~  \s^>(\p,\o')  +\s_{HF}(\p) =t^>(p)
 \ee 
\esubs

The forms of $\s^<$ and $\s^>$ stated here agree precisely with the result of second order perturbation theory, with the $\mathcal{Q}$-value described in \eq{second}.  Many other approximate results for scattering and correlation are encompassed by our formulas of \eq{simplestphi}.  We shall use the quantities, $t^>$ and $t^<$, defined here in our later calculations  for conserved quantities and entropies.

\subsection{From Boltzmann back to Maxwell (again)}
Once more we describe the result of calculating the average of an observable. Our goal is to analyze the structure of the generalized Boltzmann equation in sufficient depth so that we can approach the question of how that equation will enable us to understand the generality of the entropy concept.  We we are basically retracing Boltzmann's route in this more quantum mechanical domain. 

With this goal in mind,  we once again multiply the kinetic equation by representations of the operators in question, now $O(\p,\o,\R,T),$
and integrate over $\p,\o$.  The result is of the same general form as in the case of the  Boltzmann kinetic equation, i.e, 
\bsubs 
\be \la{O_Average}
\text{BT}_O+ \text{LHT}_O =\text{RHT}_O
\ee
Here the base term is the part of the equation of motion that would apply to non-interacting particles:
\be
\text{BT}_O= \int ~dp~O~ [g_0^{-1},g^<]
\ee
and the left hand terms is the reversible part of the motion engendered by the interaction  
\be
\text{LHT}_O= -\int ~dp~O \big\{[\s,g^<] + [\s^<,g]\big\}
\ee
while the right hand term is the irreversible, collision generated, part
\be
\text{RHT}_O= -\int ~dp~O \big\{ g^< \s^>  ~-~ g^> \s^< \big\}
\ee
\esubs

\subsubsection{The base term}
This bracket identity of \eq{BracketIdentity} immediately translates this term into a three-part expression
\bsubs
\be \la{bracket_identity_base}
\int ~ dp ~ O ~[g_0^{-1},g^<]=  \partial_T { \rho_O}^0 +\nabla_\R \cdot{ \mathbf{j}_O}^0 -{F_O}^0 
\ee
Here ${\rho_O}^0$, ${\mathbf{j}_O}^0$ and ${F_O}^0$ 
are respectively the density, current, and generalized force for the operators described by $O$ in a system without interactions among its particles. In the presence of a scalar potential, $g_0^{-1} =\o-\e-U$ so that these terms have the form: 
\be \la{rho_O}
{\rho_O}^0=\int dp ~O  ( \partial_\o g_0^{-1}) g^<(p) = \int dp~O(p) g^<(p) =<O>_T
\ee
so that we are, once again, calculating the time derivative of an averaged quantity.  The current for this quantity  is
\be \la{j_O}
{\mathbf{j}_O}^0=-\int dp ~O  ( \nabla_\p g_0^{-1}) g^<(p) =\int dp ~O  ( \nabla_\p \e(p)) g^<(p)
\ee
since the momentum derivative of the one-particle energy is the particle velocity.  Finally, the generalized force has the value.
\be \la{F_O}
{F_O} =  \int dp~  [  g_0^{-1},O] g^<(p)
\ee
The name {\em generalized force} applies especially in the case in which we are calculating conservation laws for number, moimentum and energy.  In these cases, $O$ is independent of $\R$ and $T$  and the generalized force comes directly and solely from the external potentials.\footnote{To see analysis of the situation when $g_0^{-1}$ contains vector and scalar potentials see \cite{Maciejko}[pages 30 and following].} 
\esubs
\subsubsection{The collision term}
The collision term produces the overall decay to equilibrium, and hence plays a crucial role in the analysis of entropy.  In the original entropy argument it was important to see that the conservation of particle number, energy, and momentum was reflected in the vanishing of the collision effect upon the rates of change of these quantities.  Hence we here wish to check this effect in the context of the quantum generalization.  
The collision terms in the generalized Boltzmann equation reduces to
\bea 
\text{RHT}_O &=& -\int d\p~d\o~O(\p,\o,\R, T) \nonumber \\
&~&  [ g^<(\p,\o,\R, T) \s^>(\p,\o,\R, T) -  g^>(\p,\o,\R, T) \s^<(\p,\o,\R, T) \nonumber \\
\eea
We aim to have this term vanish whenever $O$ describes one of the conserved quantities.  Notice that this result is exactly in the form that enables us to use a variational replacement equation, like \eq{VariationalReplacement}, and the generator $\psi$ of \eq{simplestpsi} to get the result, analogous to \eq{conservation}
\bea \la{rightconservation} 
\text{RHT}_O  &=& - \frac 12 \int {dp~dq~dp'~dq'}~g^<(p) ~g^<(q)~g^>(p')~g^>(q') 
\nonumber \\
&~& [O(p)+ O(q)-O(p')-O(q')]
\nonumber \\
&& \mathcal{Q}
\left(\begin{array} {ccc} p &  \rightarrow& p' \\
q &  \rightarrow& q' \\
 \end{array}\right) 
 \delta(p+q-p'-q') 
\eea   
We have left out the $\R,T$ arguments so as to simplify the writing.  For number conservation, $O=1$ so that this collision average vanishes. It equally vanishes  for the momentum and energy cases, in which $O$ is respectively $\p$ and $\o$, in virtue of the delta functions.

\subsubsection{The LHT term}
The correlation terms on the LHS of the generalized Boltzmann equation will generate correlation corrections to the behavior of entropy-like quantities in the non-equilibrium domain.     To understand the structure of these quantities, we want to check known qualitative results against the structure generated by the generalized Boltzmann equation. Specifically, we know that in fluids there are no correlation corrections to the differential particle conservation law, nor are there any to the momentum density.  How do these results emerge from our analysis? 
 
The two-body correlation term on the left hand side of the Boltzmann equation has a slightly more complex structure than the collision term on the right.  Since they both have the same source, however, their physics should be closely related. The left hand term is
\bea
\text{LHT}_O &=& -\int d\p~d\o~O(\p,\o)   \nonumber  \\
&~&
  \Big\{[\s_{HF}(\p)+ \text{Pr}  \int \frac{~d\o'}{2 \pi}\frac{  \s^>(\p,\o') -\varsigma \s^<(\p,\o')}{\o-\o'},g^<(\p,\o)] 
   \nonumber  \\
&+&[\s^<(\p,\o), 
\text{Pr} \int \frac{d\o'}{2 \pi} \frac{g^>(\p,\o') -\varsigma g^<(\p,\o')}{\o-\o'}]
\Big\}  \nonumber \\ 
\eea

Next, we simplify this expression. If $O(\p,\o)$ is independent of frequency the two terms involving a product of $\s^<$ and $g^<$ cancel each other. One can see that from simply writing the frequency integrals explicitly, and noting that the differentiations defined by the bracket all commute with the principle value integral and that the principle value integral cane be ``moved'' from one integrand to the other. Simplification for the case  of energy conservation requires a slightly more detailed argument. For this reason, we put aside discussion of the   energy conservation law.  For the other cases, once again we move the principle value integral and we are left with  
\be  \la{LHT}
\text{LHT}_O=- \int dp~O(p)    
~\big\{
 [ t^>(p),g^<(p)]
+[ -t^<(p),g^>(p)] \big\}   
\ee
with the abbreviations $t^<$ and $t^>$ defined by \eq{t<} and \eq{t>}.

As in the discussion of the base term in the  of the observable equation, \eq{bracket_identity_base}, we can split the result into three terms 
\bsubs  \la{LHTO}
\be \la{bracket_identity_left}
\text{LHT}_O=  \partial_T { \rho_O}^{LHT} +\nabla_\R \cdot{ \mathbf{j}_O}^{LHT} -\text{X}^{LHT}
\ee
They have the form: 
\be \la{rhoLHT}
{\rho_O}^{LHT}=- \int dp ~O \Big\{ ~( \partial_\o g^<(p)) ~ t^> +(\partial_\o g^>(p) )~  t^< \Big\}
\ee
\be \la{jLHT}
{\mathbf{j}_O}^{LHT}=\int dp ~O \Big\{ ( \nabla_\p g^<(p))~  t^> + ( \nabla_\p g^>(p))~  t^<\Big\}
\ee
\be \la{FLHT}
{X_O}^{LHT} = \int dp~\Big\{ [O, g^<(p)]~ t^> +[O, g^>(p)]~ t^< \Big\}
\ee
\esubs
The third contribution here, X$_O$$^{LHT}$ is quite different from the corresponding contribution in our discussion of the base term.   In that case, the third term  was directly proportional to space and time derivatives of external potentials and thus had the physical significance of a generalized force.   For the  conservations laws of number,  momentum, and energy   X$_O$$^{LHT}$ will be respectively zero, proportional to  $\nabla_\R$,  and proportional to $\partial_t$ . Thus, in the latter two cases,  this contribution will  add to the momentum current and energy density.

\subsection{Variational replacement}\la{QuantumReplacement}
Since $t^<$ and $t^>$ are variational derivatives of $\phi$, the  structures in \eq{LHS} and thus \eq{rhoLHT}, \eq{jLHT} and \eq{FLHT} are directly analogous to that found in our Boltzmann analysis of variational replacement (See \eqs{VariationalReplacement}.).  We work from the $\phi$ given in \eq{simplestphi}. Specifically,
\bsubs
\bea  \la{VAR}
\chi[O^>,O^<] =\int dp~
\big[
 O^<(p) g^<(p) t^>(p)+
O^>(p) g^>(p) t^<(p)  
\big] 
\nonumber  \\
\eea 
has as before the effect of producing a sum of terms in which each $g$ on the right hand side of \eq{simplestphi} is replaced by $Og$ so that  
\bea
\chi[O^>,O^<] &=&
  \frac1{ \pi} ~\text{Pr}~\int dp~dp'~dq~dq'  [O^<(p)+O^<(q)+O^>(p')+O^>(q')]  \nonumber  \\
&~&\frac{ g^<(p)~g^<(q)~g^>(p')~ g^>(q')} {p_0+q_0-q_0'-p_0'}   \nonumber \\
&& \mathcal{Q}
\left(\begin{array} {ccc} p &  \rightarrow& p' \\
q &  \rightarrow& q' \\
 \end{array}\right) 
 \delta(\p+\q-\p'-\q')  
  \nonumber \\
\nonumber \\
&+&\frac 12  \int ~ dp ~dp' ~[O^<(p)+O^<(p')]g^<(p) g^<(p')[\tilde{v}(0)~
+\varsigma ~\tilde{v}(\p-\p')]
  \nonumber  \\
\eea
\esubs
\subsection{Number Conservation}
One interesting thing about many-body corrections to the number conservation law  is that, in a fluid, there are no such corrections.  We therefore look back at \eqs{LHTO} for our left hand side. We set $O=1$ to represent number conservation and see what we get.

  Since the Poisson bracket of $O$ with anything is zero, the generalized force correction of \eq{FLHT} vanishes.  The current correction of \eq{jLHT} is more complex.  A comparison of \eq{jLHT} with our theorem, \eq{phitheorem} for using  the generating function says that in this case $O^>$ and $O^<$ are both $\nabla_\p$.  After an integration by parts the left hand term for the particle current then reduces to
\bea \la{phitheorem}
j^{LHT} &=&
 -\frac1{ \pi} ~\int dp~dp'~dq~dq'    \nonumber  \\
&~&\frac{ [g^<(p)~g^<(q)~g^>(p')~ g^>(q'))]} {p_0+q_0-q_0'-p_0'}  \delta(\p+\q-\p'-\q')   \nonumber \\
&& [\nabla_\p+\nabla_\q+\nabla_{\p'}+\nabla_{\q'}] \mathcal{Q}
\left(\begin{array} {ccc} p &  \rightarrow& p' \\
q &  \rightarrow& q' \\
 \end{array}\right)  
 \eea
I was surprised that this current did not vanish.   However, in a fluid, this vanishing is a consequence of Galilean invariance which precisely requires that no shift of all momentum or frequencies in the system by a fixed amount can change any scattering rate.  In symbols,
\be  \la{GalileoQuantum}
\text{if~~} h^>(p)=g^>(p+u)  \text{~~~and~~~}  h^<(p)=g^<(p+u)   \nonumber
\ee
then
\be
\phi[h^>,h^<]  = \phi[g^>,g^<] 
\ee
This result holds, for example, when $\mathcal{Q}$ is set by second order perturbation theory as in \eq{second}.
If this requirement of \eq{GalileoQuantum} on $\mathcal{Q}$ fails, as it does when we have a metal with its fixed periodic potentials, the particle current is replaced by the quasiparticle current and is thus modified by interaction effects\footnote{I am not sure what happens in a superconductor or superfluid, but in any case the state of these systems is not properly described in terms of just $g^<$ and $g^<$.}.  
The argument is the same for the particle density.   We can convert the density correction term into one proportional to the derivative of   $\mathcal{Q}$ with respect to all of its frequencies.   This too should be zero in virtue of \eq{GalileoQuantum}.  Thus, the behavior of the density conservation law is precisely what we should have expected:  For the translationally invariant system, the particle density and current have no correlation correction. 

The result in \eq{phitheorem} bears an interesting relation to the work of Baym\cite{BaymConserving}.   Baym constructed conservation laws within Green's function approximations by showing that the effect of infinitesmal application of operators like number density, number current, momentum density, etc. was to change the Green's functions according to $ G \rightarrow  G + \delta~ O ~ G $, where $\delta$ is an infinitesmal and $O$ is an operator applied to $G$.    That operator would be derivatives with respect to $\p$ and $\o$, respectively for number density and number current, $\p ~ \partial_\o$ for momentum density, etc.  Thus, as a result of  the other conservations laws, we can expect to see structures exactly like the one in \eq{phitheorem} with $\nabla_\p$ replaced by other suitable operators.

\subsection{Momentum conservation} 
Here, we are looking for a zero correlation contribution to the momentum density.  That density should be completely determined by the base term. 

As noted in the discussion of \eq{LHTO}, the contribution $X_O^{LHT}$ is proportional to a spatial gradient so that it does not contribute to the momentum density.
 
The correlation-induced change in the momentum density is then
\be \la{GrhoLHT}
{\rho_\p}^{LHT}=\int dp ~\p \Big\{ ~g^<(p) ~ \partial_\o t^> + g^>(p) ~ \partial_\o t^< \Big\} 
\ee
Once more we use the theorem of \eq{phitheorem}.  Here the operators are given by $O^>=O^<=p$. Thus we get a succession of term multiplied by $\p \partial_\o$'s for each of the scatterers.  The Hartree Foch term has no frequencies in it so it produces no correction. We are left with the term proportional to $\mathcal{Q}$.  In this term, each frequency derivative has the effect of converting the frequency denominator to the very same squared denominator. In this way, we get a result proportional to the sum of the momenta participating in the scattering.  The momentum conservation delta function then ensures that this sum is zero. So we get the expected result.

\subsection{Entropy}
For previous analyses see \cite{Nishiyama0,Nishiyama1,IKV,Friesen}.   These authors see a definition of kinetic entropy in local equilibrium situations but not otherwise.  Our conclusions will follow these precedents.
  
 Boltzmann calculated the time-dependence of  $H$ by multiplying his kinetic equation by a function of the state function, $f$ and summing over momentum space.  We follow his approach and multiply our two   kinetic equations \eq{g<} and \eq{g>} respectively by $ \ln [1/g^<(p)]+1$ and the second by $ \ln[ 1/g^>(p)]+1$ and integrate over $p$.  
Take the first equation add $- \varsigma $  times the second.  The result is a structure that takes the form
\bsubs
\be
\text{LHS}_s =\text{RHT}_s
\ee 
After a simplification based upon the conservation of particle number, the right hand side becomes
\be
\text{RHT}_s= \int dp ~ \ln (g^<(p)/g^>(p)) ~ [ g^<(p)   \s^>(p)-  g^>(p) \s^<(p) ]
\ee
The left hand side reads 
\bea \la{LHS}
\text{LHS}_s & =&  \int dp ~\Big\{[{g_0}^{-1}-\s, g^< ](-\ln g^<+1)
-\varsigma[{g_0}^{-1}-\s, g^> ](-\ln g^>+1) \nonumber \\
&+&
  [g,\s^<] (-1 +\ln g^<)-\varsigma  [g,\s^>] (-1 +\ln g^>)\Big\}
\eea
\esubs
We shall analyze these equations using expressions involving the distribution function, $f(p)$, namely
\be g^<(p) = f(p) ~ a(p)   \text{~~and~~}   g^<(p) =(1+\varsigma f(p) )~ a(p)  \ee

\subsubsection{Collision term}
We substitute the expressions of \eq{simplest<} and \eq{simplest>} into this RHT and    make use of the symmetry of \eq{InterchangeSymmetry} between $p$ and $q$  
\bea
\la{NumberConservation-integral} 
\text{RHS}_s &=& \frac 12 \int d p ~d q ~dp'~      dq'  [g^<(p) g^<(q) g^<(p') g^<(q') ]  \nonumber \\
&~&  \ln \frac{g^<(p) ~ g^<(q)~ g^>(p') ~g^>(q')}{g^>(p)~ g^>(q)~ g^<(p')~ g^<(q')}   \nonumber \\
&~& \mathcal{Q}\left(\begin{array} {ccc} p &  \rightarrow& p' \\
q &  \rightarrow& q' \\
 \end{array}\right)   \delta(p+q-p'-q')
 \nonumber \\
\eea

We assume that both $g^>$ and $g^<$ are real and non-negative.
Once again we use the detailed balance symmetry of \eq{DetailedBalance} to set up a RHS that is always  is always positive except when the system attains  local thermodynamic equilibrium, in which has $f$ in the form
\be \la{le} \frac{g^<(p)}{g^>(p)}=\frac{f(\p,\o)}{1+\varsigma f(\p,\o)} =\exp[-\b(\o-\mathbf{u} \cdot\p -\mu)]  \ee
Once again, this equilibrium involves the  local thermodynamic parameters $\b,\mathbf{u}$,and $\mu$,
If $f$ has the form required by \eq{le},  this right hand side is zero.  

Thus, we have a suitable start for constructing an entropy creation law.

\subsubsection{Entropy flow}

We make the replacements  of the $g$'s in terms of $a$ and $f$ in \eq{LHS}   
and then eliminate the logarithm of  $a$  from the equation by using \eq{identity}.  We then find
\bea \la{LHTS}
\text{LHS}_s & =&  \int dp ~\Big\{[{g_0}^{-1}-\s, a f ](1-\ln f)
-\varsigma[{g_0}^{-1}-\s, a (1+\varsigma f)  ]\big(1-\ln (1+\varsigma f) \big) \nonumber \\
&-&
  {\bf [}g,\s^<](1 - \ln f)
  +\varsigma  [g,\s^>](1 -\ln (1+\varsigma f))\Big\}
\eea 
We follow the approach of ref\cite{BotermansMalfliet}
\be \la{fs}
 \s^< =\gamma f_\s \text{~~and~~}  \s^> = \gamma(1+\varsigma f_\s ). \ee
so that we are parametrizing $\s^>$ and $\s^<$ in the same fashion as we parametrized $g^>$ and $g^<$.  If the system is in local thermodynamic equilibrium so that \eq{le} holds, then  $f_\s=f$. In that case,  we might expect to find an entropy function.  If the system has not had sufficient time to relax to this local equilibrium, we cannot know {\em a priori} whether there will be a kinetic entropy.     

Using \eq{fs}, we obtain
\bea \la{LHTS1}
\text{LHS}_s & =&  \int dp ~\Big\{[{g_0}^{-1}-\s, a f ](1-\ln f)
-\varsigma[{g_0}^{-1}-\s, a (1+\varsigma f)  ]\big(1-\ln (1+\varsigma f) \big) \nonumber \\
&-&
  {\bf [}g,\gamma f_\s ](1 - \ln f)
  +\varsigma  [g,\gamma(1+\varsigma f_\s )](1 -\ln (1+\varsigma f))\Big\}
\eea 

Since the bracket is a differential operator, we can do an integration to move the logarithms inside the brackets.  In this way, we find the relatively simple and quite useful expression in which the left hand side is a sum of two terms. the first being a local equilibrium result and the second a correction for deviation from this equilibrium 
\bsubs   \la{LHSLE}
\be  \la{LHS2}
\text{LHS}_s ={ \text{LHS}_s}^{LE}+{\text{LHS}_s}^\text{Corr}
\ee
with the local equilibrium term having the form
\bea \la{LHS3}
{ \text{LHS}_s}^{LE} & =&  \int dp ~\big\{[{g_0}^{-1}-\s,a \mathcal{S}]
-[g,\gamma \mathcal{S}]  \big\}
\nonumber \\
&~&\text{~where~} \mathcal{S} = - f (\ln f)+\varsigma (1+\varsigma f) \ln(1+\varsigma f)
\eea
and the correction having the form 
\be   \la{LHS4}
{\text{LHS}_s}^\text{Corr}=\big[g,\gamma(f_\s-f)]~  \ln  \frac {f}{1+\varsigma f}
\ee
\esubs

\subsubsection{Local equilibrium}
When the system has been left alone for a long time, it falls into a local equilibrium in which $f_\s=f$  in the broad sense that not only are the quantities equal but also their derivatives with respect to $p,\R,T$.  In that case, the correction term in \eq{LHS4} vanishes and we are left with the expression of \eq{LHS3} to define the left hand side of the putative entropy equation. 

The local equilibrium expression in  \eq{LHS3} is two terms  composed of $p$-integrals of Poisson brackets with no prefactor under the integral. Terms of this form are  perfect integrals. Thus they generate a law for the creation of entropy of the form of \eq{DefineEntropy}.  The last term will ruin this law, unless it too can be cast into the form of an integral of a bracket with no prefactor.   Although I have tried, I have not seen a way of making an expression of this form. 

Thus, in the case of local equilibrium, one can find a simple expression for the entropy as has been previously done by Nishiyama and coworkers\cite{Nishiyama0, Nishiyama1} and others\cite{Friesen,IKV}.  The entropy density is  
\bsubs \la{LE}
\be
\rho_s^\text{LE}= \int dp ~  \mathcal{S} \big\{ a\partial_\o(g_0^{-1}-\s)
-\gamma \partial_\o g   \big\}
\ee
while the entropy current has the value
\be
j_s^\text{LE}= -\int dp ~  \mathcal{S} \big\{ a\nabla_p(g_0^{-1}-\s)
-\gamma \nabla_p g   \big\}
\ee
\esubs
  (See for comparison \eq{QuasiparticleEntropyDensity} and  \eq{QuasiparticleCurrentDensity}.)
  
We conclude that entropy perfectly well exists for degenerate systems in local equilibrium in the limit in which the generalized Boltzmann equation, \eq{KB}, is valid.

\subsubsection{General result}
To treat the more general case in which we do not have local equilibrium, we return to \eq{LHTS} and compare it to the local equilibrium result of \eq{LHS3}.
We find that in this more general situation the LHS is 
\bea
LHS_s&=&\int dp~\Big\{[ {g_o}^{-1}-\s,f] a(-\ln f + \ln(1+\varsigma f)) 
\nonumber \\
&~&[ {g_o}^{-1}-\s,a] (-\varsigma +\mathcal{S})) -[g,\gamma] \varsigma
\nonumber \\
&~& -[g,\s^<]\ln f +\varsigma [g,\s^>](\ln(1+\varsigma f))
\Big\}
\nonumber \\
\eea
The first two lines combine to the first line below, while the last line is expanded to give the next two.  
\bea
LHS_s&=&\int dp~\Big\{
[ {g_o}^{-1}-\s, a \mathcal{S}]  
\nonumber \\
&~& -[g,\s^<\ln f +\s^>\ln(1+\varsigma f)] \nonumber \\
&~& -\s^<[g,\ln f ]+\varsigma\s^> [g,\ln(1+\varsigma f)]
\Big\}
\nonumber \\
\eea
Now, the first two lines provide a perfect integral, the last line does not. As before, this imperfect part vanishes in local equilibrium.  

Our result is thus in the form of \eq{DefineNoEntropy} with the entropy density, entropy current,  and correction term being
\bsubs  \la{final}
\be  
\rho_s= \int dp ~  \Big\{  \mathcal{S}a\partial_\o(g_0^{-1}-\s)
-[ \s^<\ln f +\s^>\ln(1+\varsigma f)] \partial_\o g    \Big\}
\ee
and
\be
j_s= - \int dp ~  \Big\{  \mathcal{S}a\nabla_\p(g_0^{-1}-\s)
-[ \s^<\ln f +\s^>\ln(1+\varsigma f)]   \nabla_\p g  \Big\}
\ee
so that the  law replacing the entropy production law is 
\bea \la{FinalNoEntropy} 
&~&\partial_T ~ \rho_s +\nabla_\R \cdot  \mathbf{j}_s  + 
\int ~dp~ \Big[ \frac{-\s^<}{f} +  \frac{\s^>}{1+\varsigma f}  \Big] [g,f]
\nonumber \\  &~&
= \text{RHS} _s(\R,T)~\ge~0
\eea 
\esubs
Note that in local equilibrium the two terms in \eq{FinalNoEntropy} involving $\s^>$ and $\s^<$ cancel out. Further note that $\rho_s$ and $\mathbf{j}_b$ have a different form than the ones listed for local equilibrium in \eq{LE} above.  One can jiggle the equations to change $\rho_s$ and $ \mathbf{j}_s $,     but I certainly have not been able to push the  additional term in \eq{FinalNoEntropy} into a form of a perfect integral or into something that is negative semi-definite.  Hence I conclude that I cannot define entropy density of this system.

As stated earlier, the most likely explanation is that the system shows some analog of entanglement.

\renewcommand{\theequation}{A-\arabic{equation}}
\setcounter{equation}{0}
\section*{Appendix A: $\Phi$-derivability}  
Gordon Baym\cite{BaymConserving} 
showed that one could obtain approximations with very important conservation laws as well as properties of thermodynamic consistency by using approximations that he described as $\Phi$-derivable. (See also the earlier \cite{LuttingerWard}.) These approximations used a thermodynamic Green's function, $G(1,2)$, where the numerals stand for space-time coordinates. 

  The time variables are pure imaginary in the interval $[0,-i\b]$, where $\b$ is the inverse temperature in energy units. The Green's function has  a periodicity\cite{Kubo} under replacements $t \rightarrow t-i\b$.
The reason that this interval is convenient is that the Green's functions describe thermodynamic equilibrium  at  temperature $\b^{-1}$\cite{MS}. 
The Green's functions are composed of two pieces, $G^>(1,2)$ and $G^<(1,2)$ depending upon the relative size of $it_1$ and $it_2$. As descriptors of equilibrium behavior, they only depend upon this difference variable, and not upon any absolute time variable like $T=(t_1+t_2)/2$.   Both of these functions are analytic functions of  the difference variable $t_1-t_2$,  allowing these functions may be continued to the real time axis. 

One starts by constructing $\Phi$ as defined by integrals of $G(1,2)$ and the two-body potential, $V(1,2)$.  This potential is proportional to delta functions of its time difference variable.  A typical structure that appears under the integral signs is
$$      
G(1,a)G(a,1')V(a,b) G(2,b) G(b,2')     \text{~with~} V(a,b)=\delta(t_a-t_b)v(\R_1,\R_b)
$$

In an infinitesimal range of the time variables one would get a structure like
\be
\text{Pr} \int~d\o_1~d\o_{1'} d\o_2~d\o_{2'}~ \frac{G^>(\o_1,)G^<(\o_{1'},)v() G^>(\o_{2'},) G^<(\o_{2'},)}{\o_1-\o_{1'}+\o_2-\o_{2'}}    \la{A-1}
\ee

Here I have left out space variables and chosen particular values of the signs of $it$ difference variables.  Note the lack of $i$ epsilons in the frequency denominator.  These $i\e$ specifications are only necessary when some time integration is taken to infinity and here all integrations are over finite domains.   Even the principle value sign is irrelevant because whenever there is a zero in the denominator the ``boundary condition''
\bea  &~&G^>(\o_1,)G^<(\o_{1'},) G^>(\o_2,) G^<(\o_{2'},)=G^<(\o_1,)G^>(\o_{1'},)G^<(\o_2,) G^>(\o_{2'},)\nonumber \\
\text{if}&~~& \o_1-\o_{1'}+\o_2-\o_{2'}=0\eea
ensures that there is a zero in the numerator to cancel out the zero in the denominator.  The result further simplifies if $V(1,2)$ and $G(1,2)$ are functions of spatial difference variables, as one might get with spatial periodic boundary conditions.

 Once one has put together all these integrals, one has constructed a functional, $\Phi[G^>(.),G^<(.),V(.)] $ where $.$ stands either for the space-time difference variable or for the Fourier transform variables $\p,\o$.   In addition, and most important, the functional contains a large collection of frequency denominators like the ones in \eq{A-1}.  The theory says that the variational derivative of $\Phi$ obeys
 \be
\frac{\d\Phi}{\d G(1,2)} = \S(1,2) 
 \ee
 Since both $G$ and $\S$ have a discontinuous jump when their time variables become equal to one another, one could say in more detail that
 \bsubs   \la{space-time}
 \be
\frac{\d\Phi}{\d G^>(1,2)} =-i \S^<(1,2)   \text{~~for~~} it_1<it_2
\ee
while \be
 \frac{\d\Phi}{\d G^<(1,2)} =-i \S^>(1,2)  \text{~~for~~} it_2<it_1
 \ee
 \esubs
 The factors of $-i$ are inserted for later convenience in using $\Phi$.   To be consistent with \eq{space-time}, derivatives in the $\p,\o$ representation must have the value      
 \bsubs  \la{p-omega}
 \be
\frac{\d\Phi}{\d G^>(\p,\o)} =\text{Pr} \int \frac{d\o'}{\pi}~ \frac{\S^<(\p,\o')}{\o-\o'}  
\ee
while \be
 \frac{\d\Phi}{\d G^<(\p,\o)} =-\text{Pr} \int  \frac{d\o'}{\pi} \frac{\S^>(\p.\o')} {\o-\o'}   
 \ee
 \esubs
 The frequency denominators arise from the ubiquitous denominators sitting in $\Phi$, as described above.   Since
 $$
 \text{Pr} \int \frac{d\o'}{\pi}~\frac{\exp{(-i\o't)}}{\o-\o'} = -\text{sign}(t) i
 $$
 \eq{p-omega} agrees with \eq{space-time}.
 
 Note that $\Phi$ is an extensive quantity.  One of the many space-time integrals in it gives the volume of the system $\Omega$ multiplied by the time interval $-i\b$ so that 
 $$ \la{phi-derivable}
 \Phi[G^>,G^<]= -i~\b~ \Omega ~ \phi[g^>,g^<]
 $$
 The $g$'s on the right are functions of $\p$ and $\o$ such that
 \bsubs
  \be
\frac{\d\phi}{\d g^>(\p,\o)} =\text{Pr} \int \frac{d\o'}{\pi}~ \frac{\s^<(\p,\o')}{\o-\o'}  
\ee
while \be
 \frac{\d\phi}{\d g^<(\p,\o)} =-\text{Pr} \int  \frac{d\o'}{\pi} \frac{\s^>(\p.\o')} {\o-\o'}   
 \ee
 \esubs

 \renewcommand{\theequation}{B-\arabic{equation}}
\setcounter{equation}{0}
\section*{Appendix B: Entropy Correction}
The correction to the perfect integral appearing in the final formula of the text, \eq{FinalNoEntropy} is   
\be \la{Corr}
\Delta =\int ~dp~ \Big[ \frac{-\s^<}{f} +  \frac{\s^>}{1+\varsigma f}  \Big] [g,f]
=\int ~dp~  \frac{[-\s^< g^>+\s^> g^<]}{ a f (1+\varsigma f)}   [g,f]
\ee
This expression is  second order in a gradient expansion since every square bracket encloses a expression that is zero in global equilibrium.

The final expression in \eq{Corr} is just of the right form  to use a variation of the variational replacement idea of \se{QuantumReplacement} with 
\be
-O^>(p)=O^<(p)=\frac{ [g(p),f(p)] }{ a(p) ~f(p)~[1+\varsigma f(p)]}
\ee 
Then the correction term comes out in a form based upon $\psi[g^<,g^>]$ that is
\bea
\Delta&=&\frac 12 \int  ~dp ~dq~ dp' ~dq' ~  [O^<(p)+O^<(q)-O^<(p')-O^<(q')]
 \nonumber \\
 & &\mathcal{Q}
\left(\begin{array} {ccc} p &  \rightarrow& p' \\
q &  \rightarrow& q' \\
 \end{array}\right) 
 \delta(\p+\q-\p'-\q') 
  \nonumber \\
 &~&\{ g^<(p) ~g^<(q)~ g^>(p') ~g^>(q')\}
\eea
Here, as in Boltzmann's analysis of his collision term we see that most of the expression has a simple parity under the interchange of primed and unprimed variables.  Up to the last line, everything is odd under the interchange. To make the whole expression even, we antisymmetrize the last line and find that
\bea
\Delta&=&\frac 14 \int  ~dp ~dq~ dp' ~dq' ~  [O^<(p)+O^<(q)-O^<(p')-O^<(q')]
 \nonumber \\
 & &\mathcal{Q}
\left(\begin{array} {ccc} p &  \rightarrow& p' \\
q &  \rightarrow& q' \\
 \end{array}\right) 
 \delta(\p+\q-\p'-\q') 
  \nonumber \\
 &~&\{ g^<(p) ~g^<(q)~ g^>(p') ~g^>(q')- g^<(p') ~g^<(q')~ g^>(p) ~g^>(q)\}
\eea   
I have not been able to further simplify this expression.   
    
\section*{Acknowledements}
I want to thank Gordon Baym, Gene Mazenko, David Huse, Stuart Rice, Pierre Gaspard, Carl Freed, Craig Callender, and Rudro Rana Biswas for helpful discussions. The discuss of Boltzmann's kinetic equation owes a lot to a graduate course taught by Roy Glauber that I took at Harvard.  This work was partially supported by the University of Chicago NSF-MRSEC under grant number DMR-0820054

\end{document}